\documentclass[a4paper,fleqn]{cas-dc}

\usepackage[square,numbers,sort&compress,comma]{natbib}

\usepackage{booktabs}
\usepackage{subcaption}
\usepackage{siunitx}
\usepackage[version=3]{mhchem}
\usepackage{lipsum}
\usepackage[utf8]{inputenc}

\definecolor{plot_blue}{RGB}{0,113.9850,188.9550}
\definecolor{plot_orange}{RGB}{216.7500,   82.8750,   24.9900}
\definecolor{plot_yellow}{RGB}{236.895, 176.97, 31.8750}
\definecolor{plot_violet}{RGB}{125.97, 46.92, 141.78005}
\definecolor{plot_green}{RGB}{118.83, 171.87, 47.94009}
\definecolor{plot_lightblue}{RGB}{76.755, 189.975, 237.9150}
\definecolor{plot_red}{RGB}{161.925, 19.89, 46.9200}

\newcommand{\hydrogen}{H$_2$}
\newcommand{\methane}{CH$_4$}
\newcommand{\ammonia}{NH$_3$}

\newcommand{\carbondioxide}{CO$_2$}

\newcommand{\water}{H$_2$O}
\newcommand{\nitrogen}{N$_2$}

\begin{document}
\let\WriteBookmarks\relax
\def\floatpagepagefraction{1}
\def\textpagefraction{.001}

\shorttitle{Physics-Guided Flame Speed Correlation}

\shortauthors{R. Hesse et al.}

\title [mode = title]{Physics-guided laminar flame speed correlation for methane–hydrogen–air mixtures with varying dilution}                      
%
\author[1]{Raik Hesse}[type=editor,
                        auid=000,bioid=1,
                        prefix=,
                        orcid=0000-0002-0957-5666]

\cormark[1]


\ead{r.hesse@itv.rwth-aachen.de}


\credit{Conceptualization, Methodology, Investigation, Formal analysis, Validation, Data curation, Writing – original draft, Visualization}

\affiliation[1]{organization={Institute for Combustion Technology, RWTH Aachen University},
    addressline={Templergraben 64}, 
    city={Aachen},
    postcode={52056}, 
    country={Germany}}

\author[1]{ Christian Schwenzer}[%
   suffix=,
   ]
\credit{Formal analysis, Validation, Data curation, Writing – review \& editing}

\author[1]{ Roman Glaznev}[%
   suffix=,
   ]
   
\credit{Formal analysis, Validation, Data curation, Writing – review \& editing}

\author[1]{ Florence Cameron}[%
   suffix=,
   ]

\credit{Formal analysis, Writing – review \& editing}   

\author[1]{ Heinz Pitsch}[%
   suffix=,
   ]

\credit{Writing – review \& editing, Supervision, Funding acquisition}    

\author[1]{ Joachim Beeckmann}[%
   suffix=,
   ]
\credit{Validation, Writing – review \& editing, Resources, Funding acquisition}     

\cortext[cor1]{Corresponding author}



\begin{abstract}
Fuel-flexible, low-carbon combustion systems need to accommodate methane/hydrogen mixtures with air and exhaust-gas dilution. To develop these, we require accurate and efficient correlations for laminar flame speed (LFS). In this work, we introduce a physics-guided LFS correlation that applies to burners, gas engines, and turbines. Our model uses a core-kinetic approach based on flame temperatures, an algebraic function for the equivalence ratio, and a mass-flux-based blending law. This allows for accurate predictions with any methane/hydrogen blend. We set the model parameters using one-dimensional flame simulations with C3Mech v4.0.1, chosen for its high prediction accuracy for a wide range of experimental data, including new results from our spherical combustion chamber. The new correlation provides accuracy comparable to a machine learning approach (Gaussian process regression), yet remains physically consistent, differentiable, and extrapolates well. This makes it suitable for computational fluid dynamics and control of fuel-flexible combustion systems.
\end{abstract}


\begin{highlights}
  \item Laminar flame speeds (LFS) were measured for \hydrogen{}/air and \methane{}/air mixtures. 
  \item Recent kinetic mechanisms were evaluated against these and 4000 literature targets.
  \item Analytical, physics-guided LFS correlation models dilution and \hydrogen{}/\methane{} blending. 
  \item It was benchmarked against power-law correlations and a machine-learning model (GPR). 
  \item Accurate (MAPE\,$\!<\!\qty{4}{\%}$) and robust predictions \qty{150}{bar} and \SI{1100}{\kelvin} were achieved. 
\end{highlights}


\maketitle





\section{Introduction\label{sec:introduction}} 

Decarbonization of high-temperature industrial and \linebreak[4]\mbox{power-generating} processes requires fuel-flexible combustion systems that accommodate natural gas, hydrogen, and their blends. While large-scale renewable hydrogen production is not yet established, blending hydrogen~(\hydrogen{}) with methane~(\methane{}) offers a near-term route for difficult-to-electrify sectors such as chemicals, steel, maritime, and aviation~\cite{Pitsch2024_transition}. However, \hydrogen{} and \methane{} display markedly different combustion behavior. Most notably, hydrogen-enriched mixtures have much higher laminar flame speeds, which impedes their introduction into existing combustion systems. Accurate prediction of the unstretched laminar flame speed~(LFS), $S_\mathrm{L}$, under application conditions is therefore essential for designing fuel-flexible systems. As a fundamental property linking chemical reactivity, exothermicity, and diffusivity, $S_\mathrm{L}$ enters models of flashback and quenching~\cite{Kiyamaz2022_flashback_hydrogen}, flame kernel development~\cite{Beeckmann2017CI_H2, Hesse2022experimental}, and turbulent combustion~\cite{Peters2000, Ratzke2015_turbulent_flamespeed}. 

Laminar flame speeds are measured in spherical, counterflow, flat, and Bunsen flames~\cite{Halter2005characterization, Beeckmann2019CI_flame_propagation, Bariki2020Isochoric, Wu1985determination, Hermanns2007phd, Eckart2022ch4_h2, Han2024_methane_diluted}. Methane/air mixtures have been extensively characterized. Studies include elevated temperatures and pressures relevant to gas turbines and engines~\cite{Luo_2021, Rozenchan2002, Bariki2020Isochoric, Wang2020_methane_correlation}. In contrast, most LFS studies for \hydrogen/air and \methane/\hydrogen/air have focused on atmospheric pressure~\cite{Halter2005characterization, Krejci2013hydrogen, Varea2015_hydrogen}. The high susceptibility of hydrogen flames to thermo-diffusive and hydrodynamic instabilities occurring at elevated pressure leads to wrinkled fronts and increased consumption speeds~\cite{Jomaas2007transition}. In such regimes, extensively validated detailed chemical-kinetic models remain the most reliable tools. For \methane/air, Wang et al.~\cite{Wang2020_methane_correlation} showed that well-validated mechanisms can predict engine-relevant LFS (up to \qty{30}{bar} and \SI{700}{\kelvin}) with \qty{15}{\%} accuracy.

Direct use of detailed kinetics in CFD, multi-parameter design studies, or real-time combustion control is often computationally unfeasible. Analytical LFS correlations are therefore needed that reproduce the accuracy of detailed chemistry at a fraction of the cost and can be parametrized with experiments or simulations~\cite{Wang2020_methane_correlation}. Three broad classes are widely used: empirical power-law correlations, fully data-driven surrogates, and physics-based models grounded in flame theory and asymptotics. 

Power-law correlations, originating from Metghalchi and Keck~\cite{Metghalchi1982_CF_flame_speeds}, express the laminar flame speed as a product of reference values and power-law corrections with respect to pressure, temperature, and equivalence ratio. Modern variants employ higher-order polynomial expansions for greater flexibility, for example, in engine-oriented works by Amirante et al.~\cite{Amirante_2017}, Kuppa et al.~\cite{Kuppa2018_c1_c3_alkanes_hydrogen}, and Harbi and Farooq~\cite{Harbi2020}. These correlations are computationally inexpensive and accurate within their calibration range, but they can exhibit unphysical behavior, e.g., negative flame speeds or apparent extrema, when extrapolated~\cite{Wang2020_methane_correlation}. Fully data-driven models, such as Gaussian process regressions, can also be trained using reaction-kinetic models. They map large datasets better than empirical power-law models, but extrapolate poorly.

Physics-based correlations for laminar flame speed can be derived using rate-ratio asymptotics. Williams and Peters~\cite{Williams1987asymptotic} established such expressions for stoichiometric \methane/air flames. Seshadri and Göttgens~\cite{Seshadri1991_asymptotics} extended them to lean \methane/air mixtures. Seshadri and co-workers developed related formulations for rich \methane/air and lean \hydrogen/air flames~\cite{Seshadri1998_CF_asymp_mod_rich, Seshadri2001_CF_asympt_rich_fl, Seshadri1994hydrogen}. Hermanns~\cite{Hermanns2007phd} combined these building blocks for \methane/\hydrogen/air blends and reported reasonable trends for \hydrogen{} molar fractions in the fuel stream up to \SI{50}{\percent}. These models provide deep insight into the flame structure, but are cumbersome as engineering tools, because the laminar flame speed is generally defined implicitly through inner-layer considerations. Important variables, such as the inner-layer temperature, $T_\mathrm{i}$, must be determined from nonlinear subproblems. In addition, different asymptotic formulations are required for lean, stoichiometric, and rich regions, leading to piecewise definitions that are difficult to use in design and optimization.

For these reasons, most applied work does not employ the full asymptotic formulations. Analytical expressions that are less closely guided by them are more common. The approximation proposed by Göttgens et al.~\cite{Goettgens1992_CI_approximation},
\begin{equation}    
S_\mathrm{L} = S_0\,Y_\mathrm{f,u}^m 
\exp\!\bigg(\!-\frac{\Theta_a}{T_\mathrm{i}}\bigg)
\left(\frac{T_\mathrm{u}}{T_\mathrm{i}}\right) 
\left(\frac{T_\mathrm{b}-T_\mathrm{i}}{T_\mathrm{b}-T_\mathrm{u}}\right)^n ,    
\label{eq:base_approx}
\end{equation}
has therefore seen much wider use. It appears in engine studies by Ewald~\cite{Ewald2006_level_flame}, Hann et al.~\cite{Hann2017influence, Hann2020}, Beeckmann et al.~\cite{Beeckmann2017Approx_formula_methane_air}, and Hesse et al.~\cite{Hesse2018LBVsurrogates, Hesse2022experimental}. The expression retains the asymptotic scalings with the unburned and adiabatic flame temperatures ($T_\mathrm{u}$, $T_\mathrm{b}$), the inner-layer temperature $T_\mathrm{i}$, and the unburned fuel mass fraction $Y_{\mathrm{f,u}}$. Inner-layer details are integrated into a small set of tunable parameters $S_0$, $\Theta_a$, $m$, and $n$. The expression is explicit, continuously differentiable, and inexpensive, making it a suitable basis for extensions to dilution and fuel blending.

Dilution effects can, in principle, be captured implicitly through the quantities in Eq.~\eqref{eq:base_approx}, since external diluents modify $Y_{\mathrm{f,u}}$, $T_\mathrm{b}$, and $T_\mathrm{i}$. In practice, however, both asymptotics-based and empirical correlations usually introduce an additional attenuation term that relies explicitly on a global diluent fraction~\cite{Hann2017influence, Duva2020methane_correlation, Han2024_methane_diluted}. It is often given as a linear, polynomial, or power-law function of the diluent mass fraction $Y_\mathrm{d}$. Han et al.~\cite{Han2024_methane_diluted, Han2024_ammonia_methane} recently explained the physical basis of such terms. They demonstrated that for fixed pressure and unburned temperature, the normalized flame speed decays approximately log-linearly with the inverse undiluted mass fraction $Y_\mathrm{ud}\!=\!1-Y_\mathrm{d}$ over a wide range of diluents. 

In addition to external dilution, the fuel composition strongly affects laminar flame speeds. Simple mass- or mole-fraction fuel blending laws treat $S_\mathrm{L}$ as nearly linear in composition, but for \methane/\hydrogen/air, the dependence on the hydrogen fraction is strongly nonlinear and accompanied by a pronounced shift of the peak reactivity towards hydrogen-rich mixtures. By analogy with flammability limits, several studies have adopted Le Châtelier-type blending rules for LFS, using harmonic-mean combinations of neat-fuel flame speeds weighted by fuel-side mole fractions~\cite{DiSarli2007_hydrogen_methane, Kuppa2018_c1_c3_alkanes_hydrogen}. These relations enforce the correct limits for the neat fuels and perform reasonably well under lean conditions, but lose accuracy as the hydrogen fraction grows and the equivalence-ratio location of the maximum LFS shifts~\cite{DiSarli2007_hydrogen_methane}.

To better describe nonlinear blending effects, Chen et al.~\cite{Chen2012_binary_blends} proposed a mass-flux-based blending rule for binary \methane/\hydrogen/air mixtures. Their method blends the unburned mass flux \mbox{$\dot{m}\!=\!\rho_\mathrm{u} S_\mathrm{L}$} rather than $S_\mathrm{L}$, using a three-point spline in composition anchored at the two neat fuels and one intermediate blend. This formulation is based on classical flame theory and mass continuity. It describes the growing nonlinear increase and shift of the maximum of $S_\mathrm{L}$ upon the addition of hydrogen much better than a linear or Le Châtelier mixing rule. In the present work, we adopt this mass-flux-based blending strategy as the basis for handling intermediate \methane/\hydrogen{} compositions within our physics-guided LFS framework.

Current LFS correlations can be summarized as follows: Fully asymptotic models must be solved numerically and are too complex for routine use. Empirical models rely on functional relationships and mixing rules, which can result in non-physical LFS profiles at the edges of the training dataset. Available model parameterizations are mostly tied to older reaction kinetic mechanisms and are severely limited in their operating windows. This creates a need for a compact, physics-guided correlation grounded in the latest reaction-kinetics knowledge and validated against it. In this context, the following research questions will be investigated: 
\begin{enumerate}
\item[Q1] What level of accuracy is realistically required of an analytical LFS model, and which detailed chemical kinetic mechanism provides the most consistent reference for \methane/\hydrogen/air mixtures across the relevant $p$--$T_\mathrm{u}$--$\phi$ space?  
\item[Q2] How can dilution be represented in a compact, physics-guided correlation that remains predictive over a wide range of pressures and temperatures?   
\item[Q3] How can the effects of \hydrogen{}-enrichment on the laminar flame speed of \methane{}/air mixtures be accurately modeled for the entire blending range?    
\item[Q4] Can the new LFS correlation achieve accuracy equivalent to data-driven models, such as Gaussian process regression, while providing better extrapolation performance?
\end{enumerate}


\section{Methods\label{sec:methods}}

\begin{figure}
     \centering
    \small
    \includegraphics[trim={0cm 0cm 0cm 0.1cm},clip,width=\linewidth]{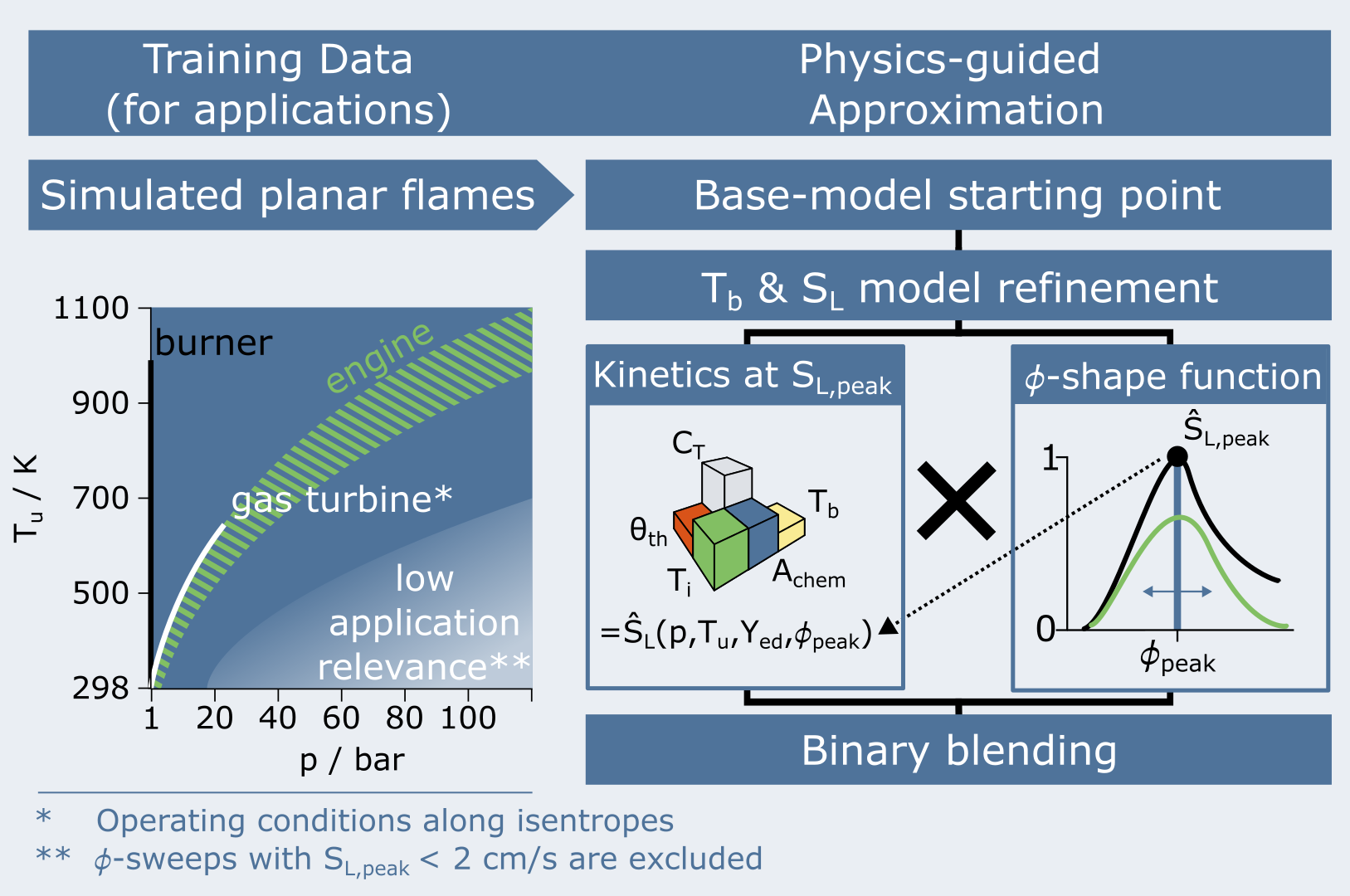}
    \caption{Overview of the LFS modeling framework.}
    \label{fig:methodology}
\end{figure}

This section describes the workflow for developing and evaluating the proposed laminar flame speed model, as outlined in Fig.~\ref{fig:methodology}. We first assemble a database of laminar flame speeds for \methane/\hydrogen/air mixtures by combining literature data with new experiments. Then, we use three recent detailed chemical kinetic mechanisms to calculate laminar flame speeds in premixed planar flames. Next, we compare measured and simulated flame speeds to select the most consistent kinetic mechanism for our application scenarios, cf. Fig.~\ref{fig:methodology}~(left). This mechanism is then used to compute laminar flame speeds for a large matrix of conditions, serving as training data for the new physics-guided correlation. After a detailed description of the proposed model, we compare it with simulated flame speeds and literature correlations retrained on the same data set. Additionally, we use Gaussian process regression as a data-driven approach, trained on the same dataset.

\subsection{Laminar flame speed data\label{subsec:literature_datasets}}

To compare chemical kinetic mechanisms and quantify the uncertainty of $S_\mathrm{L}$ predictions, we assembled a literature database of 3977 measured laminar flames. These flames were tested in flat, counterflow, Bunsen, and spherical configurations. The data include \methane/air, \methane/\hydrogen/air, and \hydrogen/air mixtures, covering the ranges summarized in Table~\ref{tab:flame_speed_summary}. About \SI{67}{\percent} of the data are for \methane/air, \SI{13}{\percent} for \methane/\hydrogen/air, and \SI{20}{\percent} for \hydrogen/air mixtures.

For \methane{}/air, comprehensive data in the literature exist for fresh gas temperatures up to approximately \SI{800}{\kelvin} and for pressures up to \qty{30}{bar}. These conditions allow kinetic mechanisms to be validated for actual burner, gas turbine, and engine conditions. In contrast, measurements of laminar flame speeds for \methane{}/\hydrogen{}/air mixtures are largely limited to atmospheric pressure and moderate preheating temperatures. Data on \hydrogen{}/air mixtures at elevated pressures remain scarce. Only a few have been collected using the same setup and data reduction method. Additional measurements at elevated pressures, using the same setup and data reduction, can be very helpful for evaluating trends and discrepancies in chemical kinetic models. 

\begin{table}[h]
    \centering
    \footnotesize
    \caption{Summary of LFS data from the literature, previous studies in the RWTH facility, and the present study.\label{tab:flame_speed_summary}}
    \begin{tabular}{lllccl}
    \toprule
    Fuel & Origin & $T_\mathrm{u}$ [K] & $p$ [bar] & $\phi$ [-] & Ref.\\
    \midrule
    CH$_4$ & Lit. & 298-803 & 1-70 & 0.5–1.7 & 
    \begin{minipage}[t]{0.19\linewidth} 
        \cite{Gu2000lbv, Rozenchan2002, Dong2002_Egol_lbv_piv, Liao2004naturalgas_markstein, Farrell2004lbv, Bosschaart2004hydrocarbons, Liao2004naturalgas, Takizawa2005cf, qin2005measurements, Halter2005characterization, LAWES2005, Law2006combustionphysics, Huang2006LBV_H2, Ogami2006_lbv_methane, Hermanns2007phd, Chen2007dimethylethermethane, tahtouh2009measurement, Hu2009CH4_H2_air_study, Halter2010, wang2010laminar, Veloo2010comparative, Kochar2010laminar, Mazas2010flamespeedch4, hermanns2010temperature, Park_2011, Chen2011_CF_extraction_flame, Mazas2011watervapormethane, lowry2011laminar, Dirrenberger2011laminar_naturalgas, Wang2012ozonemethane, Varea2012, Goswami2013, Beeckmann2014Lbv_alcohols, Zahedi_2014, Hu2014methane_co2, Hu2015ignitiondelay, Baloo2016isooctanemethane, Cai2016oxygen_enriched_methane, Nonaka2016biogas, Wu2016lbv_pressure, Khan2017, Nilsson2017laminar_alkane, Okafor2018methaneammonia, Wang2018lbv_ch4_dme_h2, Hinton2018vessel, Konnov2018comprehensive, Okafor2019methaneammonia, Turner2019methane, Han2019, Varghese2019methane, Mohammad2019dme_methane, Bradley2019hydrocarbons, Hashimoto2020, Chen2020methane_co2, Wang2020oxygen_enriched_methane_and_co2, Duva2020methane_correlation, Duva2020methane, lubrano2020comparative, Luo_2021, Duva2021two_stage_combust_methane, Khan2021methane_co2, Eckart2022ch4_h2} 
    \end{minipage}\\
    & RWTH & 298 & 1 & 0.6-1.35 & \cite{Beeckmann2019CI_flame_propagation}\\
    & RWTH & 298-373 & 1-20 & 0.6-1.35 & Present study\\
    \midrule
    \begin{minipage}[t]{0.03\linewidth}CH$_4$/ H$_2$\end{minipage} & Lit. & 298-373 & 1-5 & 0.6–2.2 & 
    \begin{minipage}[t]{0.19\linewidth} 
        \cite{Halter2005characterization, Huang2006LBV_H2, Hermanns2007phd, Hu2009CH4_H2_air_study, ECM_2017_Kruse_H2_CH4_mix, Wang2018lbv_ch4_dme_h2, Eckart2022ch4_h2} 
    \end{minipage}\\
     & RWTH & 298-373 & 1-3 & 0.6, 0.8 & \cite{ECM_2017_Kruse_H2_CH4_mix, PrataliMaffei2025C3MechV4}\\
    \midrule
    H$_2$ & Lit. & 295-443 & 1-10 & 0.2–7.2 & 
    \begin{minipage}[t]{0.19\linewidth} 
        \cite{Kwon2001Premixed_H2_flames, Li2004updated, Conaire2004modelinghydrogen, Dahoe2005_laminar, Verhelst2005hydrogen, Huang2006LBV_H2, Bradley2007hydrogen, Hu2009CH4_H2_air_study, Dong_2009_laminarflamespeed, Pareja2010hydrogen, Pareja2011LBV_H2, Hong2011, Kuznetsov2012flammabilitylimits, kumar2013experimental, Krejci2013hydrogen, dayma2014peculiar, Varga2014_CI_optimization, Varea2015_hydrogen, Konnov2015excitedhydrogen, ichikawa2015laminar, Konnov2018comprehensive,  Darocha2019ammonia, Konnov2019h2model, Wang2021ammoniahydrogen, Chen2022hydrogen_oxygen} 
    \end{minipage}\\
    & RWTH & 298-423 & 1-3 & 0.5-2.0 & \cite{ICDERS_2017_H2_FlameSpeeds, Beeckmann2017CI_H2}\\
    & RWTH & 298-423 & 1-5 & 1.5-2.5 & Present study\\
    \bottomrule
\end{tabular}
\end{table}

\begin{figure}
	\centering
	\small
    \begin{subfigure}[b]{.24\linewidth}
        \includegraphics[trim={0.0cm 0cm 0.0cm 0cm},clip,width=\linewidth]{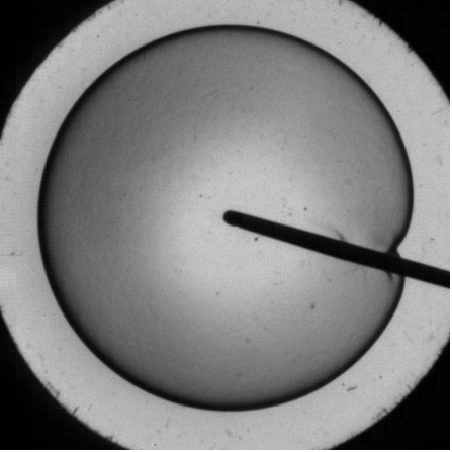} 
        \caption{\methane{}/air \qty{1.013}{bar}}
        \label{fig:flame_image_a}
    \end{subfigure}
    \begin{subfigure}[b]{.24\linewidth}
        \includegraphics[trim={0.0cm 0cm 0.0cm 0cm},clip,width=\linewidth]{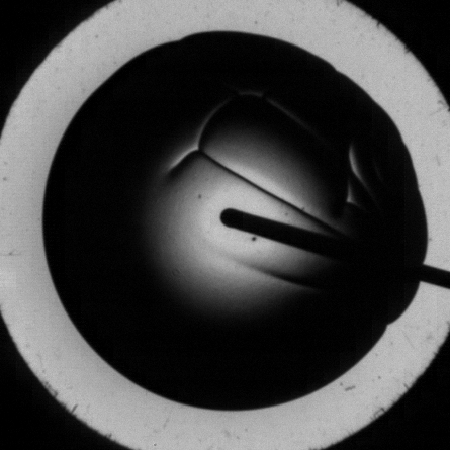} 
        \caption{\methane{}/air \qty{7.5}{bar}}
        \label{fig:flame_image_b}
    \end{subfigure}
    \begin{subfigure}[b]{.24\linewidth}
        \includegraphics[trim={0.0cm 0cm 0.0cm 0cm},clip,width=\linewidth]{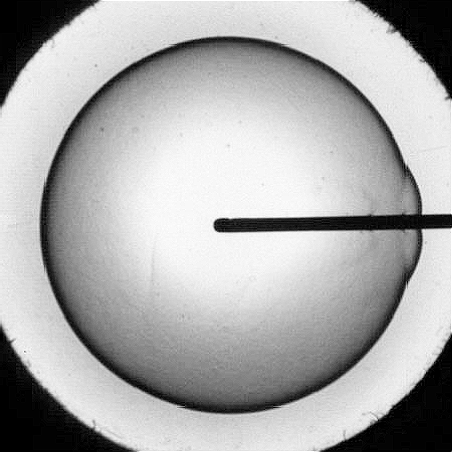} 
        \caption{\hydrogen{}/air \qty{1.013}{bar} \cite{Beeckmann2017CI_H2}}
        \label{fig:flame_image_c}
    \end{subfigure}
    \begin{subfigure}[b]{.24\linewidth}
        \includegraphics[trim={0.0cm 0cm 0.0cm 0cm},clip,width=\linewidth]{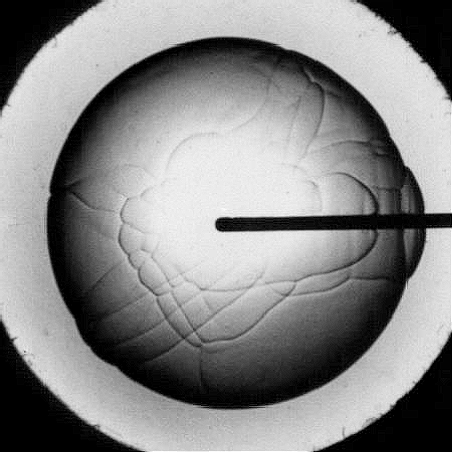} 
        \caption{\hydrogen{}/air \qty{2.0}{bar} \cite{Beeckmann2017CI_H2}}
        \label{fig:flame_image_d}
    \end{subfigure}
	\caption{Spherical flame images for $\phi\!= 1.0$ mixtures, preheat temperatures of \SI{373}{\kelvin}, and a radius of \SI{18}{\milli\meter} recorded in the RWTH setup: Example flames from the present study and Beeckmann et al.~\cite{Beeckmann2017CI_H2}.}
	\label{fig:flame_images}
\end{figure}

To close the above-mentioned gaps at elevated pressures and to establish a uniform validation basis for \methane{}/\hydrogen{}/air mixtures, we conducted new spherical combustion chamber experiments. The same spherical combustion chamber, at RWTH Aachen University, and data evaluation methodology were used as in our previous studies on \methane{}/\hydrogen{}/air mixtures by Beeckmann et al.~\cite{ICDERS_2017_H2_FlameSpeeds, Beeckmann2017CI_H2,  Beeckmann2019CI_flame_propagation}, and Bariki et al.~\cite{Bariki2020Isochoric}. We refer the reader to these works for detailed descriptions of the setup, data reduction, and uncertainty analysis. A brief summary is provided here, with more detailed information in the Supplementary Material. 

The interior of the combustion chamber, which measures \SI{100}{\milli\meter} in diameter, is observed using a linear schlieren setup through two optical access points. The unstretched laminar flame speeds and the Markstein lengths were determined from spherically smooth flames during their quasi-isobaric propagation using standard methods for stretch extrapolation and for conversion into laminar flame speeds relative to the unburned gas using density ratio corrections. 

The operating conditions, covered by the new and previous measurements in the RWTH facility, are also shown in Table~\ref{tab:flame_speed_summary}, and example flame images are depicted in Fig.~\ref{fig:flame_images}. The new data extend the existing dataset to higher pressures for \methane/air and \hydrogen/air and provide matched measurements for \methane/\hydrogen/air blends. \methane{}/air mixtures were measured up to \qty{20}{bar}. For \hydrogen{}/air mixtures, thermodiffusive instabilities enhance flame propagation, particularly for fuel-lean mixtures. This effect is further amplified by hydrodynamic instabilities at elevated pressures~\cite{Beeckmann2017CI_H2}. Thus, stable flame conditions were limited to \qty{2}{bar} for lean mixtures. For rich mixtures, the limit increased to \qty{5}{bar}. This consistent dataset is particularly valuable because it enables comparison of detailed kinetic mechanisms and analytical correlations without inferring differences in experimental setups and flame configurations.

\subsection{Chemical kinetic mechanism comparison\label{subsec:simulation}} 

Premixed, steady, one-dimensional planar flames are simulated with FlameMaster~\cite{Pitsch1998fm}. Transport is simulated using mixture-averaged properties with non-unity Lewis numbers and Soret effects, and the flame structure is resolved with an adaptive mesh.

Three chemical-kinetic mechanisms are considered:
\begin{itemize}
    \item ITVMech by Langer et al.~\cite{Langer2022pah_mech}, 
    \item CRECKMech by Bagheri et al.~\cite{Bagheri2020creck_model}, 
    \item C3Mech v4.0.1 by Pratali Maffei et al.~\cite{PrataliMaffei2025C3MechV4}. 
\end{itemize}

\begin{figure*}[tb]
    \centering
    \small
    %
    %
    \begin{subfigure}[b]{.336\linewidth}
        \includegraphics[trim={0.0cm 0cm 0.0cm 0cm},clip,width=\linewidth]{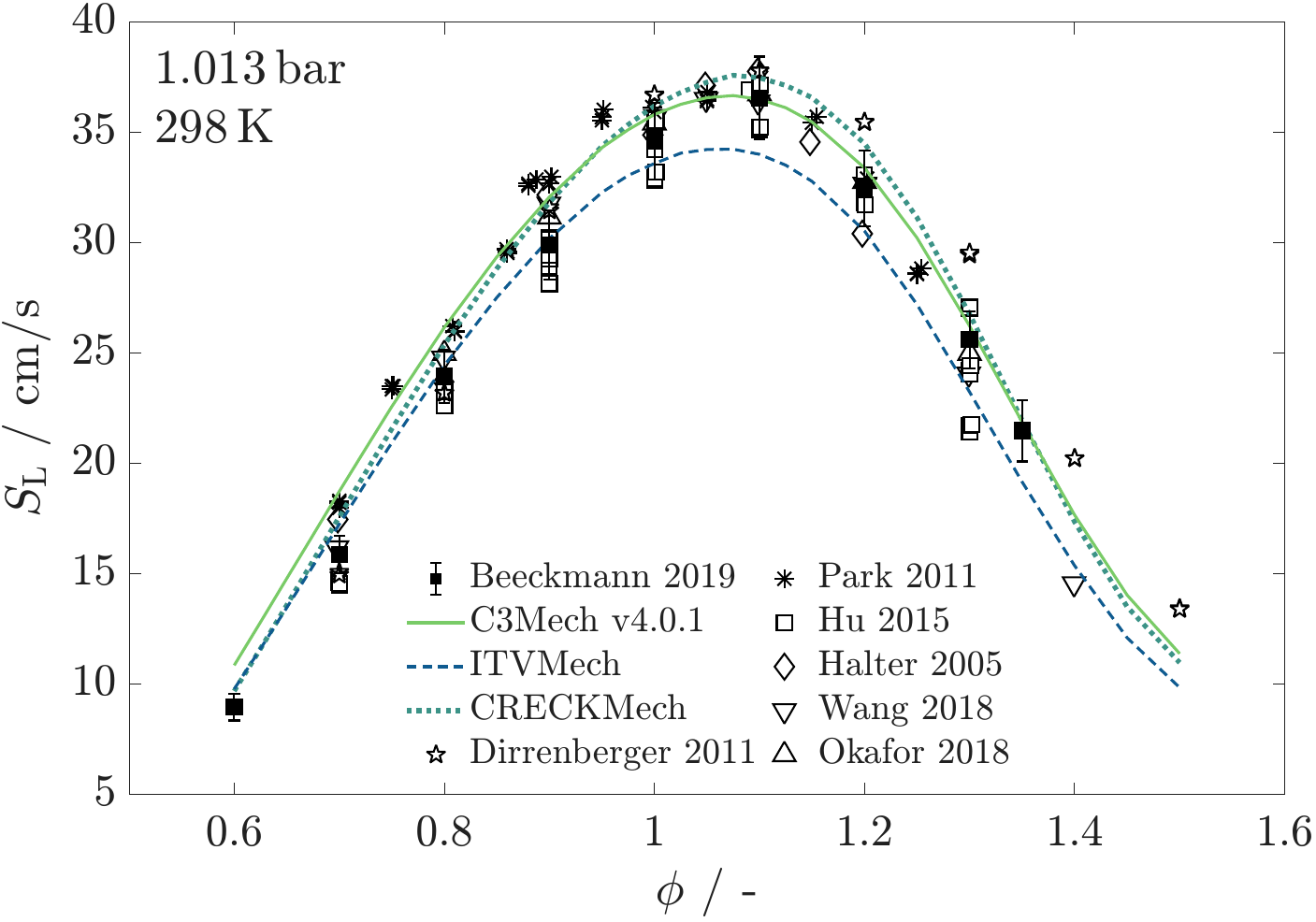}
        \caption{\methane{}/air}
        \label{fig:SL_CH4_sim_vs_exp_a}
    \end{subfigure}
    \begin{subfigure}[b]{.325\linewidth}
        \includegraphics[trim={1.0cm 0cm -0.5cm 0cm},clip,width=\linewidth]{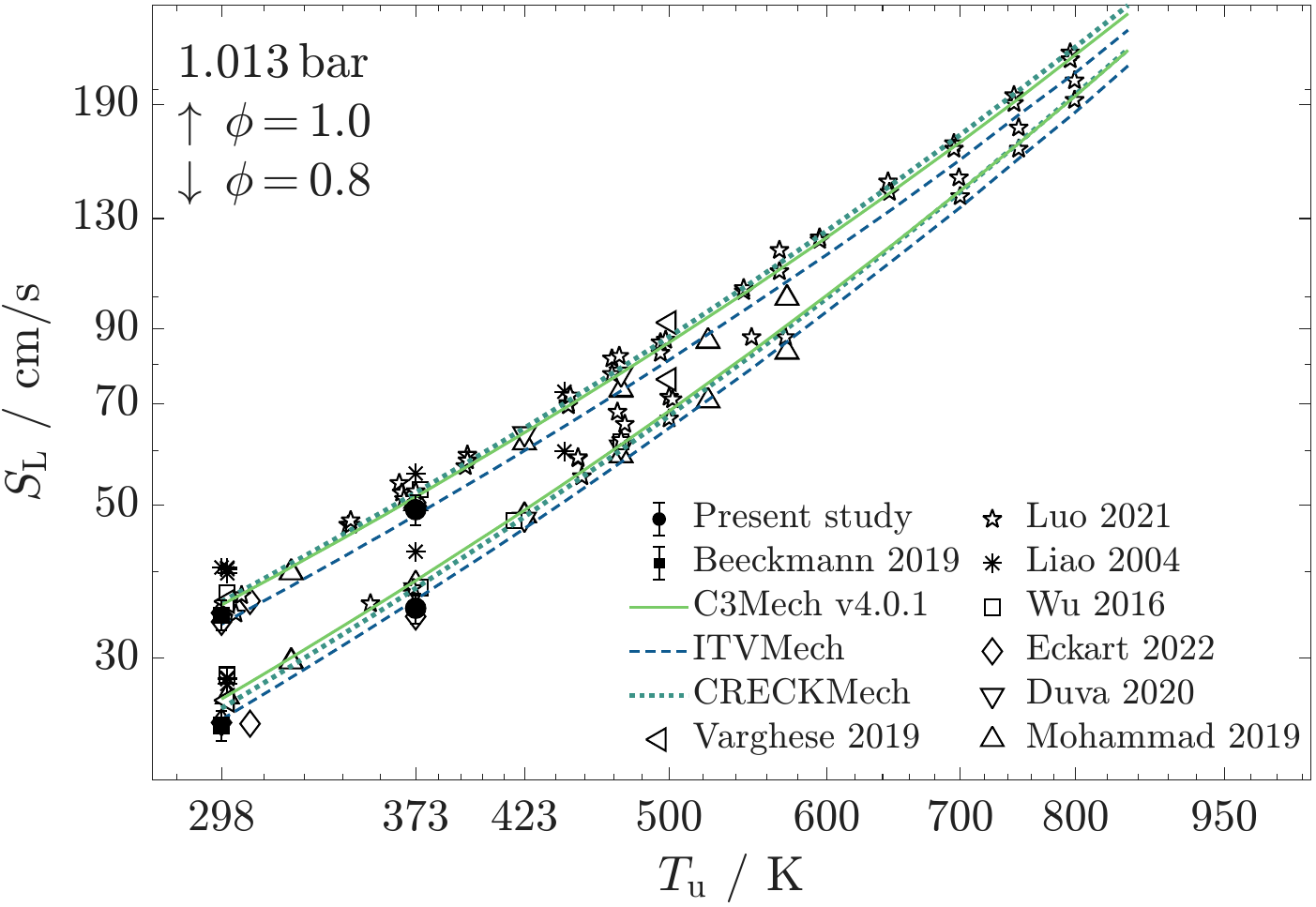}
        \caption{\methane{}/air}
        \label{fig:SL_CH4_sim_vs_exp_b}
    \end{subfigure}
    \begin{subfigure}[b]{.32\linewidth}
        \includegraphics[trim={1.0cm 0cm -0.5cm 0cm},clip,width=\linewidth]{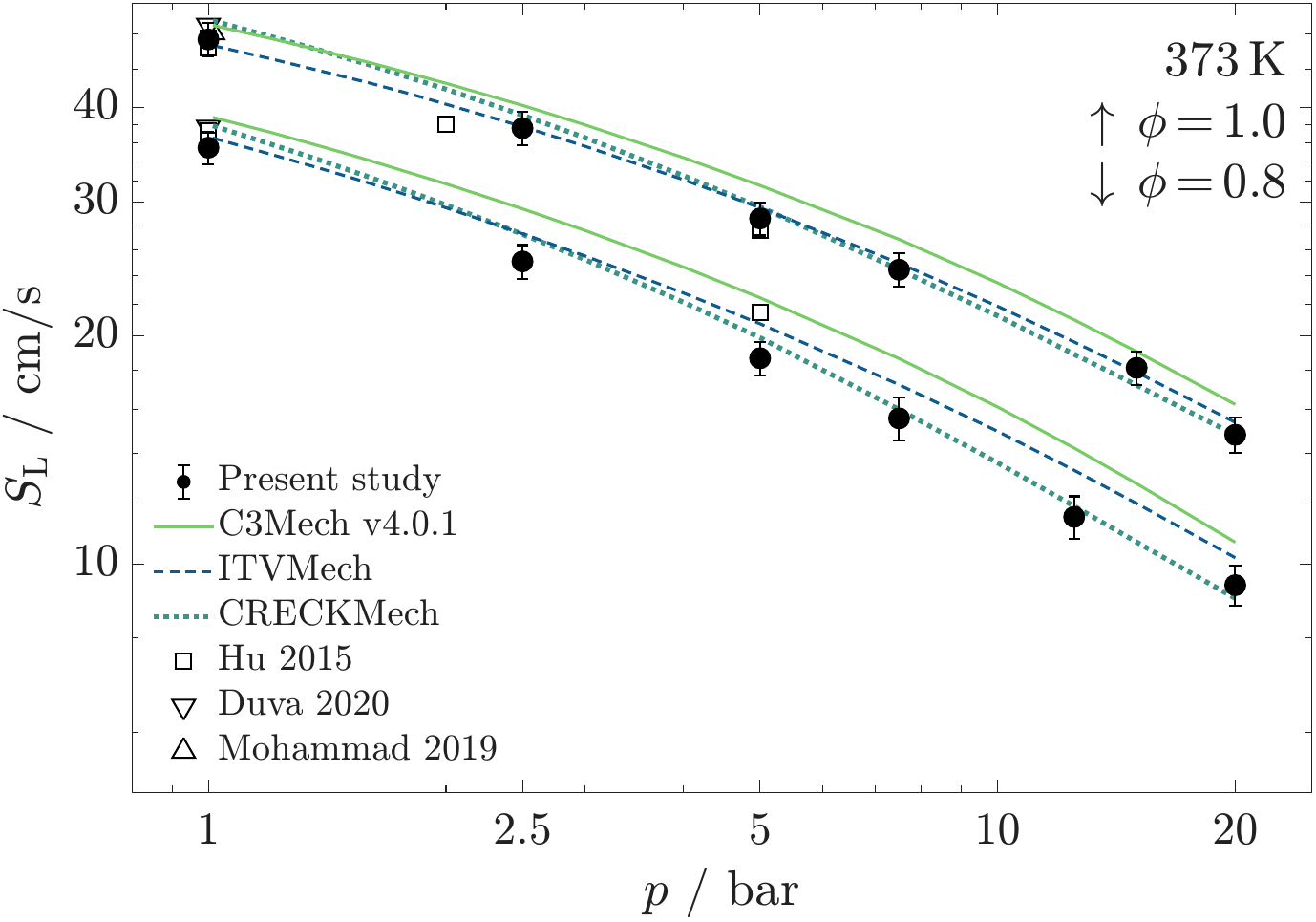} 
        \caption{\methane{}/air}
        \label{fig:SL_CH4_sim_vs_exp_c}
    \end{subfigure}\\[5pt]
    %
    %
    \begin{subfigure}[b]{.336\linewidth}
        \includegraphics[trim={0.0cm 0cm 0.0cm 0cm},clip,width=\linewidth]{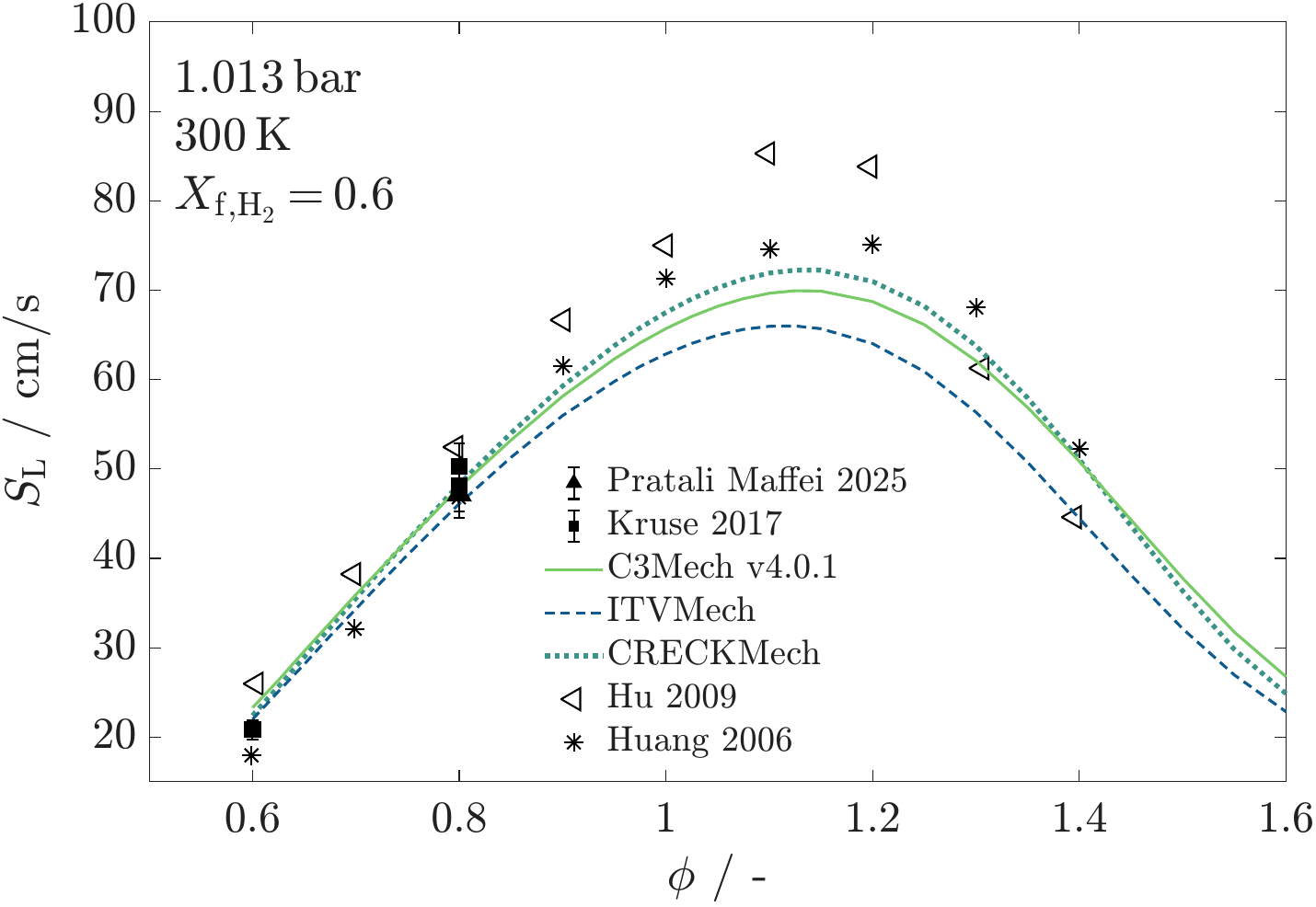}
        \caption{\methane{}/\hydrogen{}/air $X_{\mathrm{f,H}_2} = 0.6$}
        \label{fig:SL_CH4H2_sim_vs_exp_d}
    \end{subfigure}
    \begin{subfigure}[b]{.32\linewidth}
        \includegraphics[trim={1.0cm 0cm -0.5cm 0cm},clip,width=\linewidth]{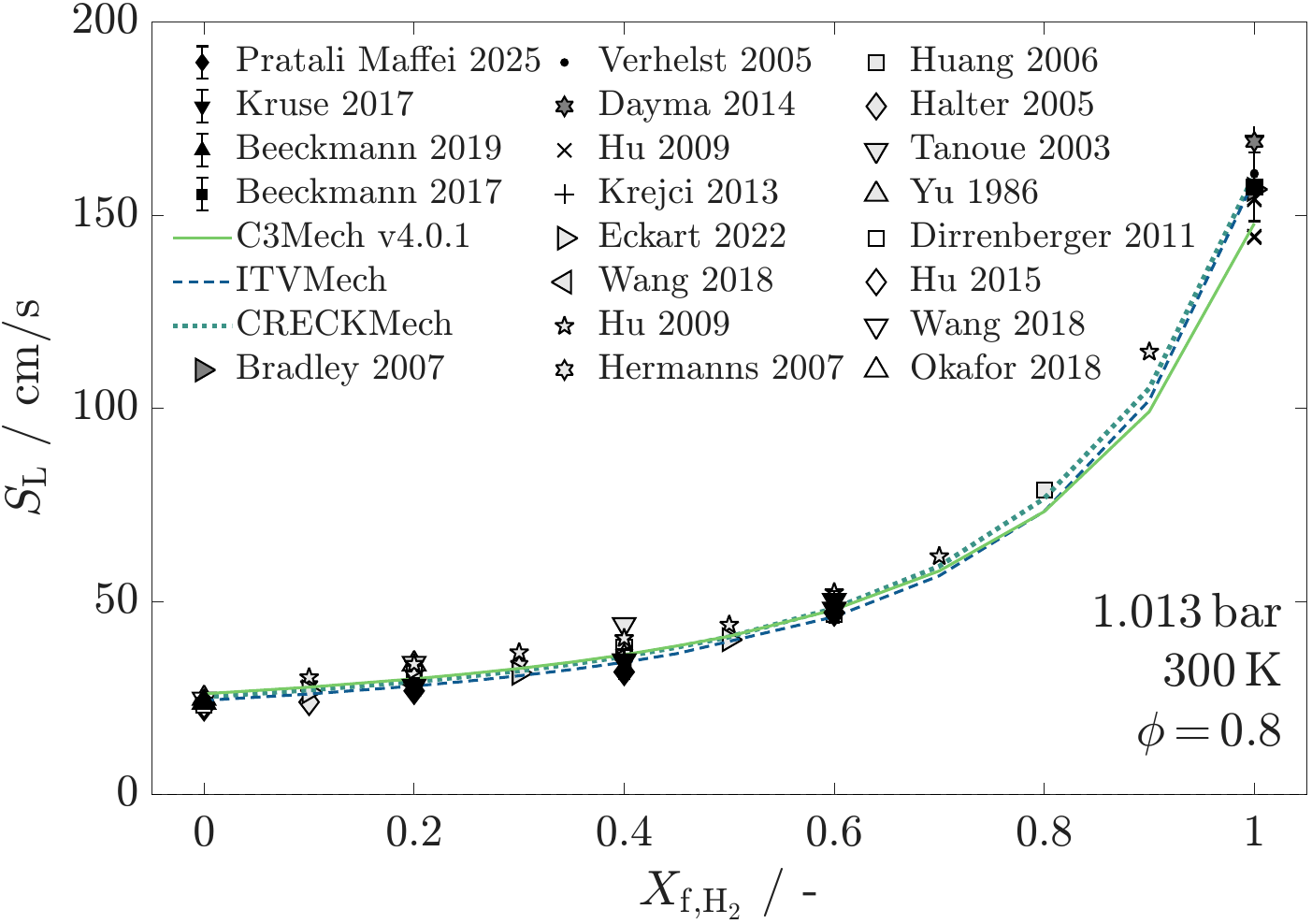}
        \caption{\methane{}/\hydrogen{}/air}
        \label{fig:SL_CH4H2_sim_vs_exp_e}
    \end{subfigure}
    \begin{subfigure}[b]{.32\linewidth}
        \includegraphics[trim={1.0cm 0cm -0.5cm 0cm},clip,width=\linewidth]{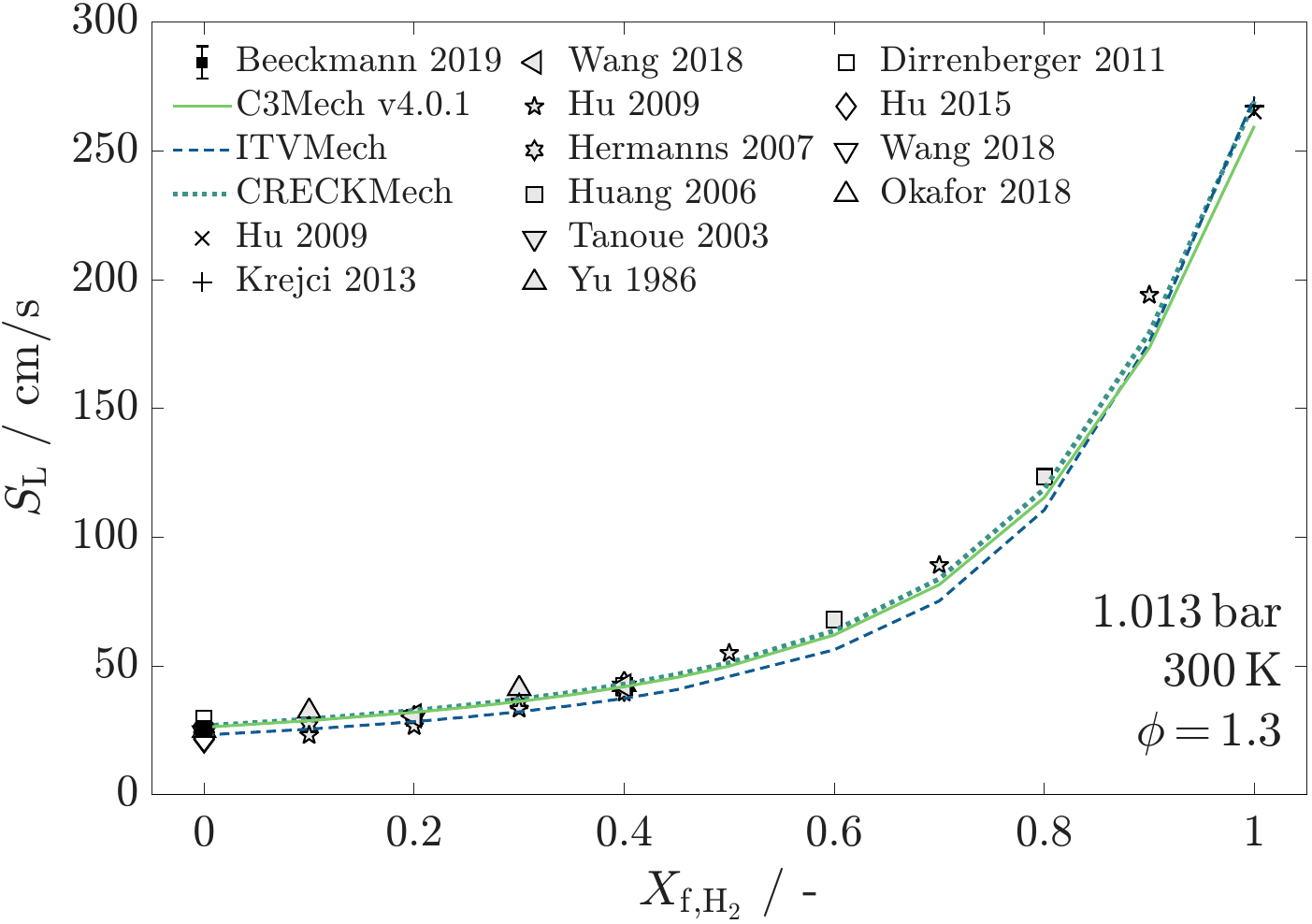} 
        \caption{\methane{}/\hydrogen{}/air}
        \label{fig:SL_CH4H2_sim_vs_exp_f}
    \end{subfigure}\\[5pt]
    %
    %
    \begin{subfigure}[b]{.336\linewidth}
        \includegraphics[trim={0.0cm 0cm -0.4cm 0cm},clip,width=\linewidth]{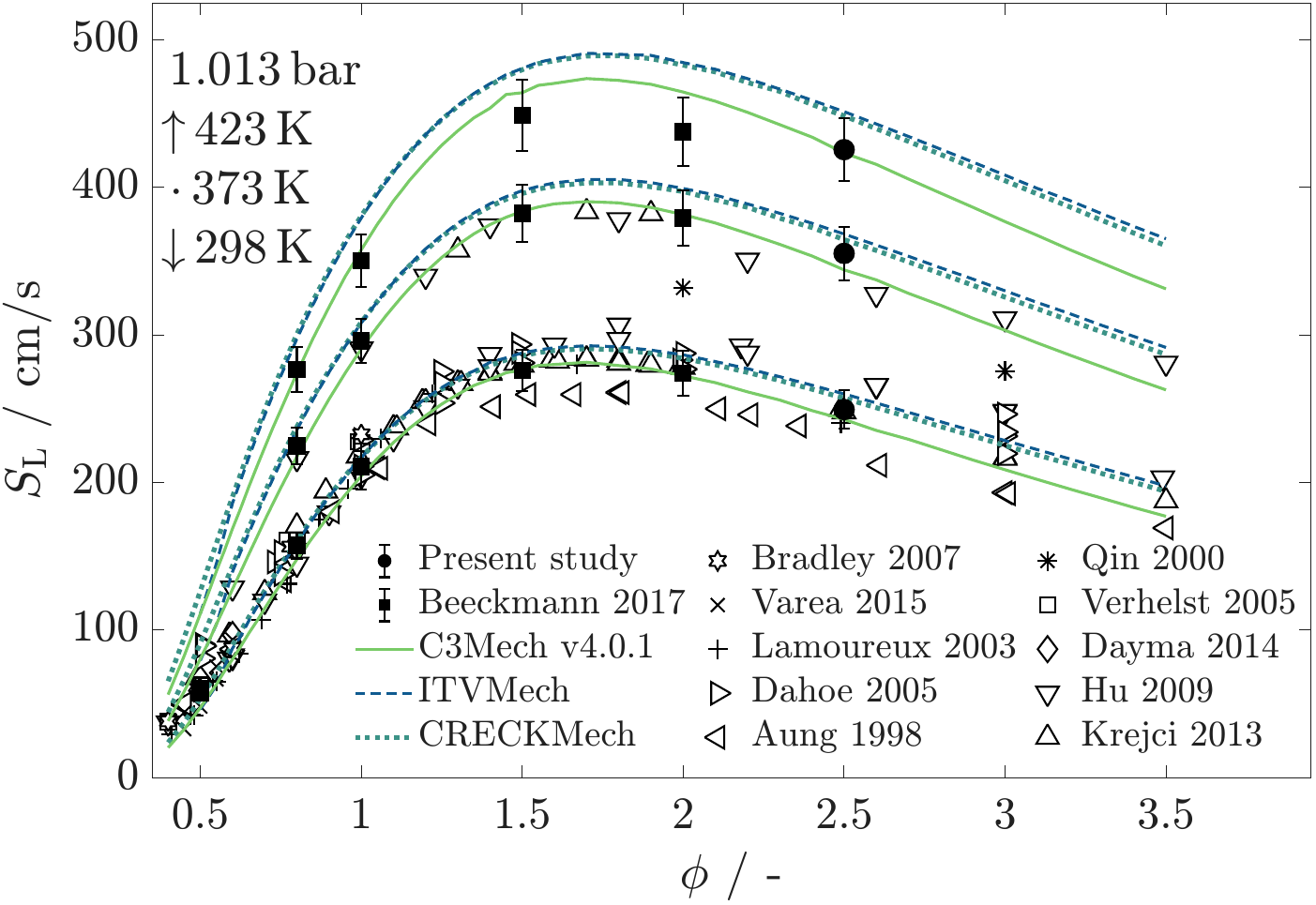}
        \caption{\hydrogen{}/air}
        \label{fig:SL_H2_sim_vs_exp_g}
    \end{subfigure}
    \begin{subfigure}[b]{.314\linewidth}
        \includegraphics[trim={1.0cm 0cm -0.4cm 0cm},clip,width=\linewidth]{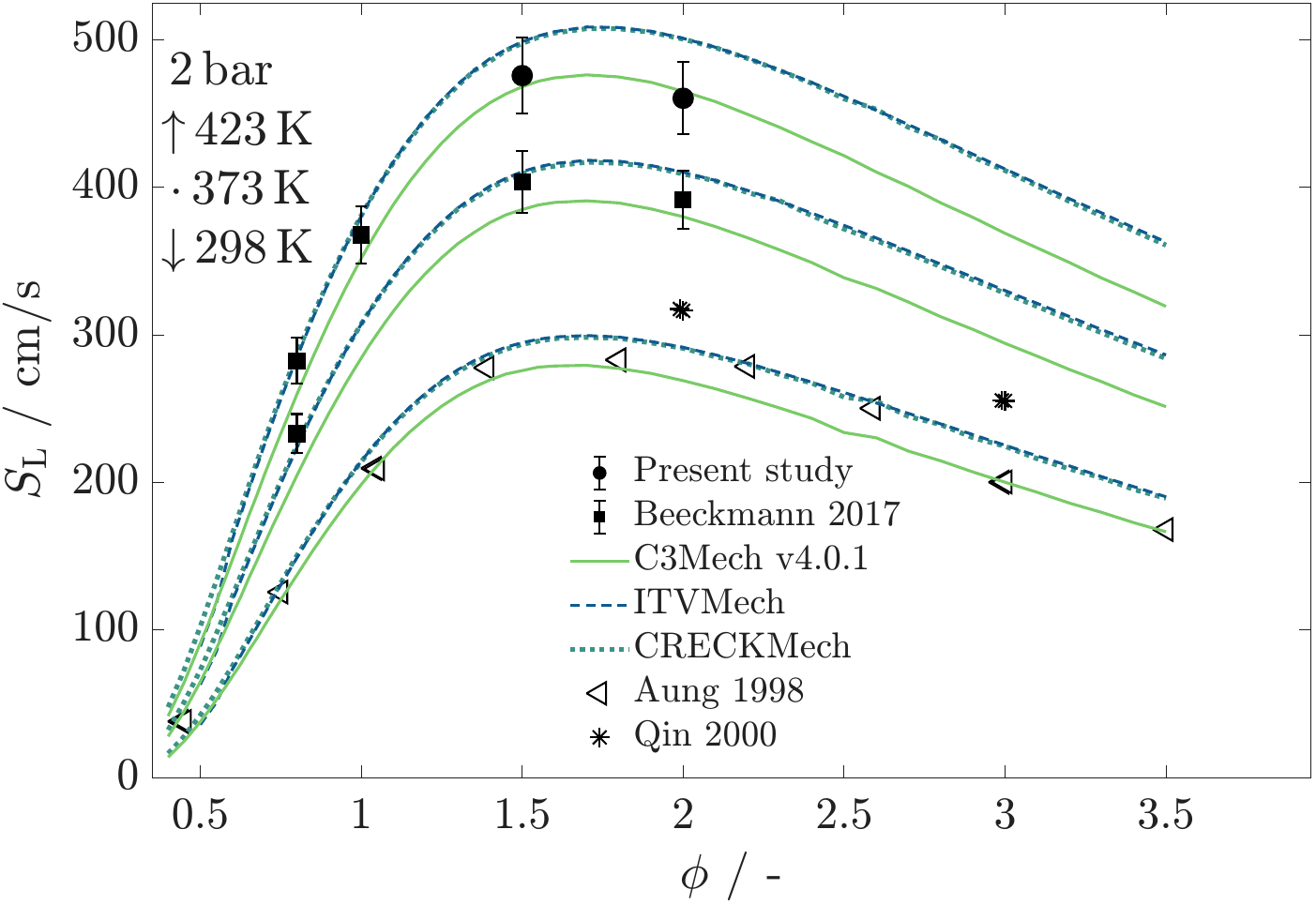}
        \caption{\hydrogen{}/air}
        \label{fig:SL_H2_sim_vs_exp_h}
    \end{subfigure}
    \begin{subfigure}[b]{.314\linewidth}
        \includegraphics[trim={1.0cm 0cm 0.0cm 0cm},clip,width=\linewidth]{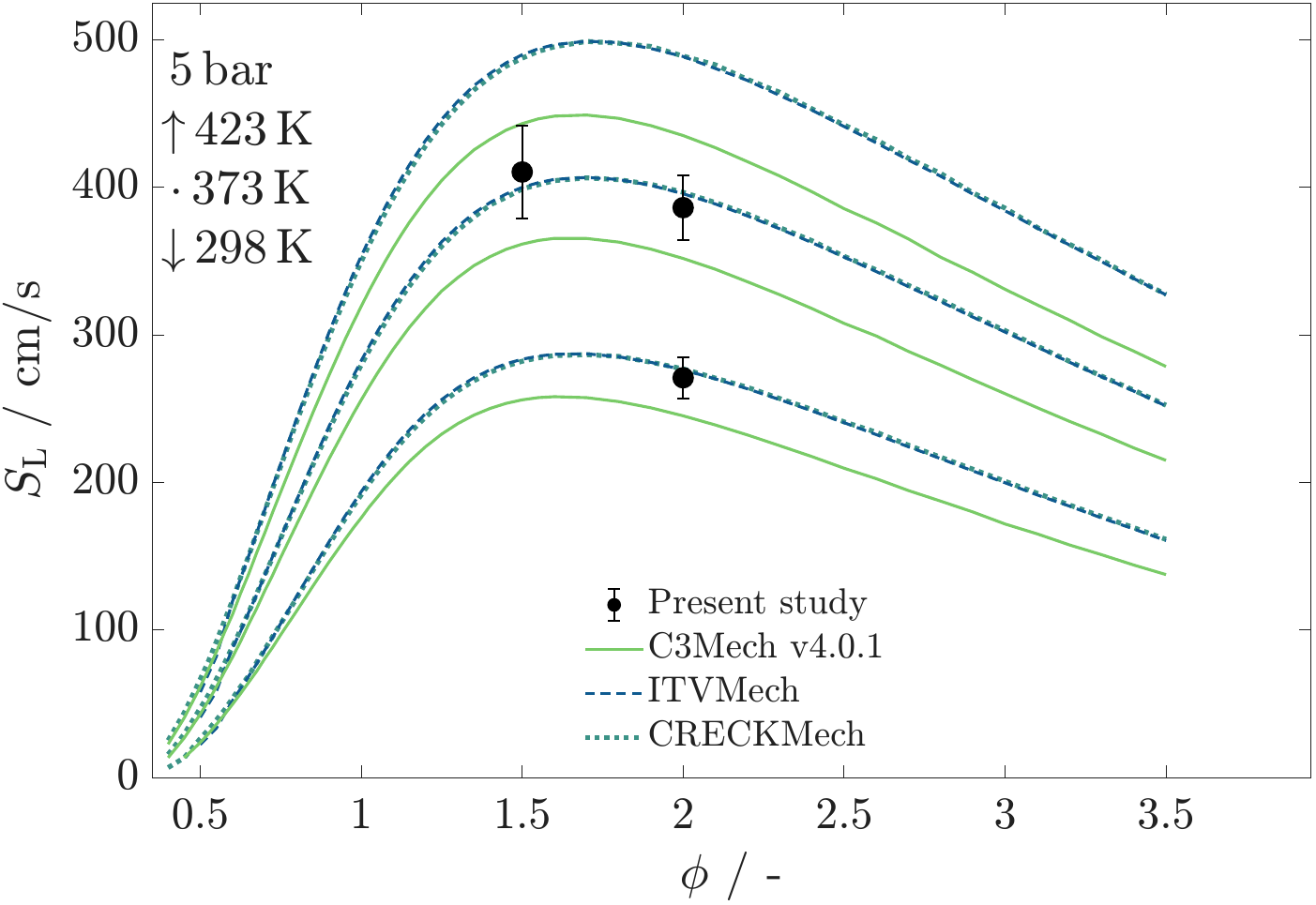} 
        \caption{\hydrogen{}/air}
        \label{fig:SL_H2_sim_vs_exp_i}
    \end{subfigure}\\[5pt]
    \caption{Comparison of chemical-kinetic mechanisms with selected LFS experiments for (a)–(c) \methane/air~\cite{Beeckmann2019CI_flame_propagation, Hesse2019simulation,Dirrenberger2011laminar_naturalgas, Park_2011, Hu2015ignitiondelay, Halter2005characterization, Wang2018lbv_ch4_dme_h2, Okafor2018methaneammonia, Varghese2019methane, Luo_2021, Liao2004naturalgas, Wu2016lbv_pressure, Eckart2022ch4_h2, Duva2020methane, Mohammad2019dme_methane}, (d)–(f) \methane/\hydrogen/air~\cite{PrataliMaffei2025C3MechV4, ECM_2017_Kruse_H2_CH4_mix, Beeckmann2019CI_flame_propagation, ICDERS_2017_H2_FlameSpeeds, Hu2009CH4_H2_air_study, Huang2006LBV_H2, Bradley2007hydrogen, Varea2015_hydrogen, Aung1998Pressure_effect_H2, Verhelst2005hydrogen, dayma2014peculiar, Hermanns2007phd, Tanoue2003JSME, Krejci2013hydrogen, Yu1986_methane_hydrogen, Wang2018lbv_ch4_dme_h2, Okafor2018methaneammonia}, and (g)–(i) \hydrogen/air mixtures~\cite{Hesse2019simulation, ICDERS_2017_H2_FlameSpeeds, Bradley2007hydrogen, Varea2015_hydrogen, Lamoureux2003bomb, Dahoe2005_laminar, Aung1998Pressure_effect_H2, Qin2000LBV_H2, Verhelst2005hydrogen, dayma2014peculiar, Hu2009CH4_H2_air_study, Krejci2013hydrogen}. Panel titles indicate $T_\mathrm{u}$, $p$, $\phi$, and $X_{\mathrm{f,H}_2}$.}
    \label{fig:SL_H2_sim_vs_exp}
\end{figure*}

The mechanisms are compared to selected \methane/air, \methane/\hydrogen/air, and \hydrogen/air LFS experiments, including the new elevated-pressure measurements. Figures~\ref{fig:SL_CH4_sim_vs_exp_a}-\ref{fig:SL_H2_sim_vs_exp_i} illustrate representative comparisons for variations in $\phi$, $T_\mathrm{u}$, $p$, and $X_{\mathrm{f,H}_2}$. For \methane/air at standard atmospheric conditions, typical experimental scatter is about \qty{5}{\%} near the peak LFS and up to \qty{10}{\%} on the rich and lean flanks~\cite{Chen2015accuracy}, and all three mechanisms reproduce these data within comparable bounds. Similar behavior is observed at elevated temperatures and pressures, and for \hydrogen{}-containing mixtures, wherever experimental data exist, although high-pressure LFS data for \hydrogen/air remain sparse.

To rank the mechanisms, we use the coefficient of determination $R^2$ and the mean absolute percentage error (MAPE), $f$,
\begin{equation}
    R^2 = 1 - 
\frac{\sum_{j=1}^N \bigl(\hat y_j - y_j\bigr)^2}{\sum_{j=1}^N \bigl(y_j - \overline{y}\bigr)^2},
\qquad
    f = \frac{1}{N}\sum_{j=1}^N
    \left|\frac{\hat y_j - y_j}{y_j}\right|,
\end{equation}
where $y_j$ and $\overline{y}$ are the measured LFS and the corresponding mean, respectively. $\hat y_j$ denotes the simulated LFSs. $R^2$ emphasizes conditions with fast absolute LFS and variance, whereas $f$ assigns relatively higher weight to slow flames. Reporting both metrics allows us to assess accuracy spanning the full range of conditions.

\begin{figure}
    \centering
    \small
    \includegraphics[trim={0.0cm -3mm 0.0cm 0cm},clip,width=\linewidth]{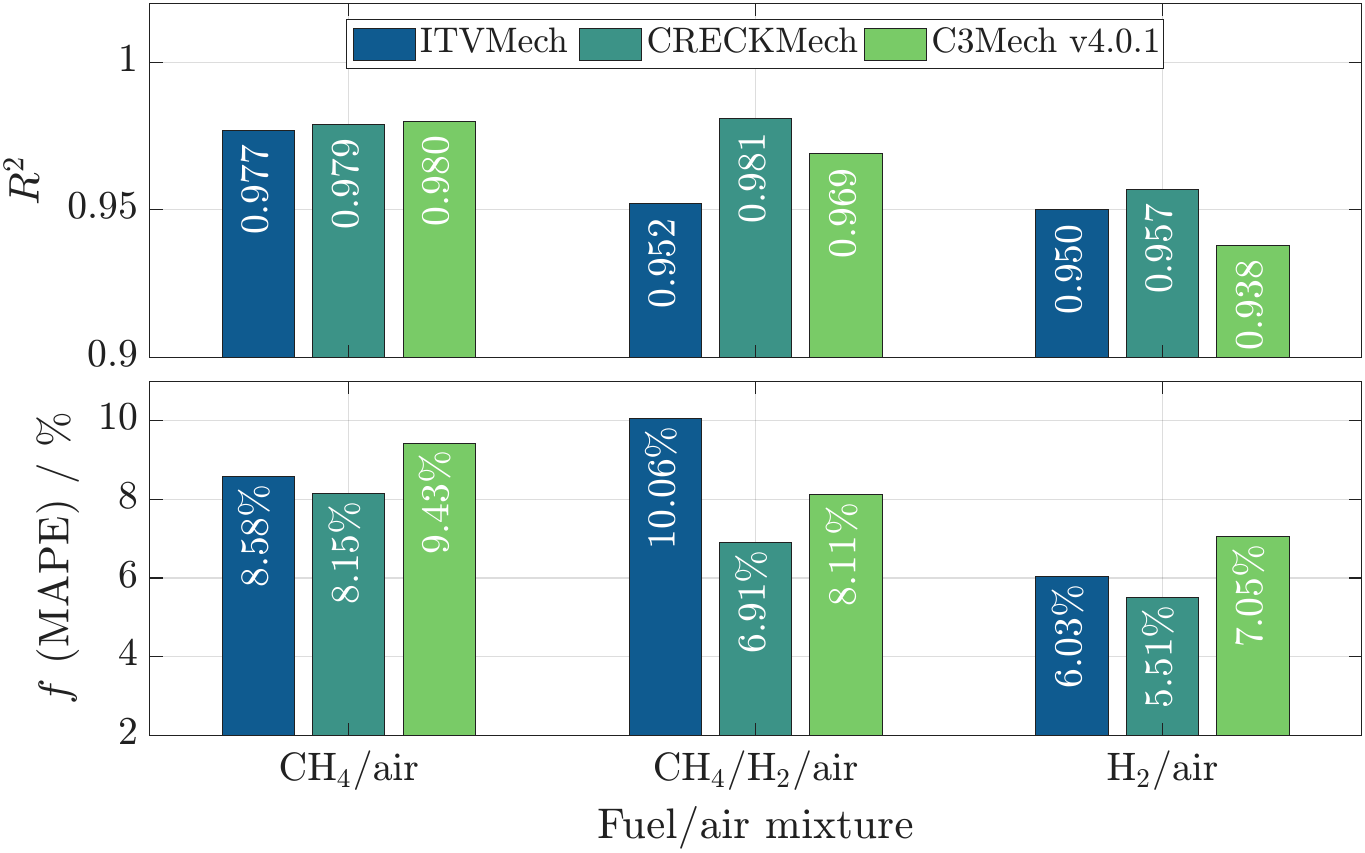} 
    \caption{Comparison of chemical-kinetic mechanisms with LFS experiments using coefficients of determination, $R^2$, and mean absolute percentage errors, f~(MAPE).}
    \label{fig:SL_kinetic_benchmark}
\end{figure}

Figure~\ref{fig:SL_kinetic_benchmark} summarizes the resulting scores. All three mechanisms achieve high $R^2$ values for \methane/air mixtures, with minor reductions when hydrogen is present. The MAPE score tends to be larger for \methane{} than for \hydrogen{}, which can be attributed to higher experimental uncertainties in slow, stretch-affected, and radiation-cooled \methane{} flames. In faster \hydrogen{} flames, measurement accuracy is primarily limited by temporal resolution, often due to limitations in the camera's acquisition rate. The remaining differences between mechanisms are similar to the variation in experimental data. Thus, the best-performing mechanisms set an empirical upper bound for LFS model expectations. This suggests a target accuracy for the proposed correlation of $R^2 \gtrsim 0.98$ and $f \lesssim 8\,\%$.

While the CRECK mechanism performs slightly better on the specific LFS benchmark used here, C3Mech v4.0.1 has been validated against a broader set of experimental targets and operating conditions and includes recent updates to hydrogen chemistry, reported by Klippenstein et al.~\cite{Klippenstein2022_HO2}. We therefore use C3Mech v4.0.1 as the reference mechanism for generating the training data.

\subsection{Simulation matrix\label{subsec:matrix}}

The training data are generated from planar-flame simulations. These simulations use isentropes that span pressures from \qty{1.013}{bar} to \qty{150}{bar} and unburned-gas temperatures from \SI{298}{\kelvin} to \SI{1100}{\kelvin}. They represent low- and high-load operating conditions in spark-ignition engines, gas turbines, and atmospheric or preheated burners. The fuel-air equivalence ratio $\phi$ ranges from 0.6 to 1.6 for \methane{}/air and from 0.2 to 4.0 for \hydrogen/air. For \methane/\hydrogen/air blends, discrete \hydrogen{}-fractions are considered for \mbox{$X_{\mathrm{f,H}_2}\!=\!0.4$} and 0.8. 

External dilution is represented by an exhaust-gas recirculation surrogate based on the findings of Duva et al.~\cite{Duva2020methane, Duva2022HydrogenMarkstein}, who showed that only combined dilution by \carbondioxide{}, \water{}, and \nitrogen{} can reproduce the effect of realistic combustion residuals on LFS. The simulation matrix covers external dilution levels $Y_\mathrm{ed}$ from 0 to 30~\% of the unburned mass fraction using a composition that depends on the hydrogen fraction in the fuel (Tab.~\ref{tab:exhaust_surrogate}). Pure \carbondioxide{} dilution, for example, would overpredict the reduction in LFS compared with realistic residual-gas dilution.

\begin{table}[tb]
    \caption{Surrogate external dilution mixtures for varying \hydrogen{} fractions.\label{tab:exhaust_surrogate}}
    \footnotesize
    \centering
    \begin{tabular}{llllll}
        \toprule
        $X_{\mathrm{f,H}_2}$ & $Y_{\mathrm{f,H}_2}$ & $Y_{\mathrm{ed,N}_2}$  & $Y_{\mathrm{ed,H}_2\mathrm{O}}$ & $Y_{\mathrm{ed,CO}_2}$ & $W_\mathrm{ed}$    \\
        mol/mol & w/w   & w/w   & w/w   & w/w   & g/mol \\
        \midrule
        0   & 0     & 0.737 & 0.123 & 0.140 & 27.53\\
        0.1 & 0.014 & 0.738 & 0.126 & 0.136 & 27.44\\
        0.2 & 0.030 & 0.739 & 0.130 & 0.131 & 27.34\\
        0.3 & 0.051 & 0.740 & 0.135 & 0.125 & 27.22\\
        0.4 & 0.077 & 0.741 & 0.141 & 0.118 & 27.07\\
        0.5 & 0.112 & 0.743 & 0.148 & 0.110 & 26.88\\
        0.6 & 0.159 & 0.745 & 0.157 & 0.099 & 26.66\\
        0.7 & 0.227 & 0.747 & 0.168 & 0.085 & 26.37\\
        0.8 & 0.335 & 0.750 & 0.184 & 0.066 & 25.98\\
        0.9 & 0.531 & 0.753 & 0.207 & 0.040 & 25.46\\
        0.95& 0.705 & 0.754 & 0.226 & 0.020 & 25.07\\
        1.0 & 1.000 & 0.756 & 0.244 & 0.000 & 24.68\\
        \bottomrule
    \end{tabular}
\end{table}

From each simulation, we extract three quantities: the adiabatic flame temperature, the unstretched laminar flame speed, and the inner-layer temperature defined as the temperature at the location of maximum temperature gradient $\partial T/\partial x\vert_\mathrm{max}$~\cite{Goettgens1992_CI_approximation}. The latter characterizes the balance between chain-branching and chain-terminating reactions in the reaction zone and is a key input to correlations derived from rate-ratio asymptotics.

\subsection{Model calibration and validation\label{subsec:fitting_approach}} 

Parameters of the physics-guided approximation are obtained by nonlinear least-squares fits to the simulated $T_\mathrm{b}$, $T_\mathrm{i}$, and $S_\mathrm{L}$ data. To improve numerical conditioning and interpretability, we normalize selected inputs, for example,
\begin{align}
\tilde T &= \frac{T_\mathrm{u}}{T_\mathrm{i,ref}},\qquad
\tilde p = \ln\Bigl(\frac{p}{p_\mathrm{ref}}\Bigr),\nonumber\\
\tilde \phi_j &= \ln\Bigl(\frac{\phi}{\phi_{j,\mathrm{ref}}}\Bigr), \quad j\in \{T,S\},
\end{align}
with $T_\mathrm{i,ref}$ and $p_\mathrm{ref}$ chosen near the geometric medians of the training data and $\phi_{j,\mathrm{ref}}$ denoting the equivalence ratios at maximum $T_\mathrm{b}$ and $S_\mathrm{L}$.

We employ weighted least squares on the training set $\mathcal{D}=\{(\phi_j,p_j, T_{\mathrm{u},j},Y_{\mathrm{ed},j},y_j)\}$ to solve the minimization problem for the parameters $\theta$,
\begin{equation}
\min_{\theta}\;
\sum_{j\in\mathcal{D}} w_j\,
\bigl[\hat y\,(\phi_j,p_j,T_{\mathrm{u},j},Y_{\mathrm{ed},j};\theta) - y_j\bigr]^2,
\end{equation}
where $\hat y_j$ denote the predictions and $ y_j$ the training data for $T_{\mathrm{b},j}$ or $S_{\mathrm{L},j}$. To avoid bias from uneven sampling in $\phi$, we use $\phi$-balanced weights that give comparable influence to lean, near-stoichiometric, and rich conditions. Then, residuals are constructed by combining linear and logarithmic weighted residuals for $S_\mathrm{L}$, so that slow and fast flames are suitably represented by the fit. In addition, residuals in $T_\mathrm{i}$ are included with a minor weight to improve inner-layer temperature predictions. All optimizations are carried out in \textsc{Matlab 2025a} using \texttt{lsqcurvefit} with constrained parameters to promote physically meaningful parameter values. A multi-start strategy with jittered initial guesses is used to mitigate the problem of finding only local minima. Model accuracy for $T_\mathrm{b}$, $T_\mathrm{i}$, and $S_\mathrm{L}$ is quantified by $R^2$ and MAPE metrics introduced in Sec.~\ref{subsec:simulation}. Additionally, cluster-based 5-fold cross-validation in $T_\mathrm{u}, p, Y_\mathrm{ed}$, and $\phi$ dimensions is carried out to assess fit quality and extrapolation capability. 

For the retraining of literature correlations, we used the same approach. However, we did not perform cross-validation, since the models remain unmodified and were most likely tested for overfitting by the original research authors. The example machine-learning method, Gaussian process regression (GPR), was selected for its better predictive performance, as evaluated against various methods using \textsc{Matlab}’s Statistics and Machine Learning Toolbox. Hyperparameters are determined by maximizing the marginal likelihood, as is common for Gaussian processes. The performance is evaluated using random 5-fold cross-validation to compute $R^2$ and MAPE.


\section{Physics-guided Correlation\label{sec:approximation_formula}} 

As sketched in Fig.~\ref{fig:methodology}, the proposed $S_\mathrm{L}$ correlation is built through physics-guided steps. First, the adiabatic flame temperature model is refined. Then, the laminar flame speed model is extended based on the approximation formula of Göttgens et al.~\cite{Goettgens1992_CI_approximation}. Model functions are separated into a reaction kinetic term, evaluated at the maximum laminar flame speed equivalence ratio, and an algebraic term that modulates the lean-to-rich fuel/air equivalence ratio effect. Finally, a mass-flux-based blending rule merges model parameterizations of neat fuels and supporting blends to continuously calculate laminar flame speeds across blend ratios. The following subsections detail each component and list parameter values. For convenience, the parameter sets for all blends and a Python model implementation are provided in the Supplementary Material.

\subsection{Adiabatic flame temperature model\label{subsec:adiabatic_flame_temperature_model}} 

As a baseline, we consider the low-order polynomial in $\phi$ with a linear dependence on $T_\mathrm{u}$ proposed by Müller et al.~\cite{Mueller1997_approximation}, here extended by a linear scaling with the external-diluent mass fraction $Y_\mathrm{ed}$,
\begin{align}
T_\mathrm{b,poly}
&= b_T\,T_\mathrm{u}
 + \sum_{k=0}^{3} b_{\phi,k} \phi^k +  b_{Y}\,Y_\mathrm{ed},
\label{eq:poly}
\end{align}
with fitted coefficients $b_T$, $b_{\phi,k}$, and $b_Y$. While simple and flexible, this purely algebraic form does not capture the correct lean/rich asymptotes and can exhibit unphysical behavior when extrapolated beyond the calibration range.

To obtain a more robust, physics-guided surrogate, we instead write
\begin{equation}
T_\mathrm{b} \;=\;
T_\mathrm{u} \;+\; \Delta T,
\end{equation}
and model the adiabatic temperature rise $\Delta T$ as
\begin{align}
\Delta T
&= A(T_\mathrm{u})\;
\frac{\bigl(1-\xi_Y\,Y_\mathrm{ed}\bigr)\,S_{\phi_T}(\phi)}
     {1 \;+\; \alpha_0\,S_{\phi_T}(\phi) \;+\; \tau_Y\,Y_\mathrm{ed}},
\label{eq:core}
\end{align}
with a linearly varying amplitude
\begin{equation}
A \;=\; A_0 \;+\; A_T\,T_\mathrm{u}.
\end{equation}
This structure represents a typical adiabatic energy balance: $\Delta T\!\sim\!Q_\text{net}/c_{p,\text{mix}}$. The numerator is an effective net heat-release term, and the denominator is a mixture heat capacity.

In Eq.~\eqref{eq:core}, the factor $S_{\phi_T}(\phi)$ is a normalized shape function that captures how the available heat release varies with the fuel-air composition. The term $1-\xi_Y Y_\mathrm{ed}$ accounts for the reduction in net exothermicity when external diluents replace reactants. $A(T_\mathrm{u})$ reflects the dependence of the maximum temperature rise on the initial temperature through both kinetics and thermal heat capacities. The mixture heat capacity in the denominator increases with $S_{\phi_T}$ and $Y_\mathrm{ed}$, which, in turn, reduces $\Delta T$.

The dependence on equivalence ratio is captured by $S_{\phi_T}$, evaluated at 
$\tilde\phi_T = \ln(\phi/\phi_{T,\mathrm{ref}})$, where $\phi_{T,\mathrm{ref}}$ denotes the equivalence ratio at maximum adiabatic flame temperature for a given fuel blend. In practice, $\phi_{T,\mathrm{ref}}$ is obtained from the detailed-chemistry data as the mean peak location over all training conditions at $Y_\mathrm{ed}=0$, and the corresponding scatter is small. We use a normalized two-parameter shape function that can be written in exponential form as
\begin{align}
S_{\phi_T}
&=\;
2^{1/q_T}\,
\exp\bigl(-m_r\,\tilde \phi_T\bigr)\,\nonumber\\
&\cdot\Bigl[1 + \exp\bigl(-(1+m_r)\,q_T\,\tilde \phi_T\bigr)\Bigr]^{-1/q_T},
\label{eq:shape_fct}
\end{align}
with fitted coefficients \mbox{$q_T\!>\!0$} and \mbox{$m_r\!\in\!(0,1]$}. By construction, \mbox{$S_{\phi_T}(0)\!=\!1$}, so that $\Delta T$ attains its maximum at \mbox{$\phi\!=\!\phi_{T,\mathrm{ref}}$}. For lean mixtures, $S_{\phi_T}$ increases almost linearly with $\phi$, while for rich mixtures it decreases proportionally to $\phi^{-m_r}$. Thus, $m_r$ controls the slope for fuel-rich mixtures and $q_T$ controls the roundness at the peak. In combination with the parameters $A_0$, $A_T$, $\alpha_0$, $\tau_Y$, $\xi_Y$, and $\phi_{T,\mathrm{ref}}$, this provides a compact and flexible description of the adiabatic flame temperature as a function of $\phi$, $T_\mathrm{u}$, and $Y_\mathrm{ed}$.

\subsection{Adiabatic flame temperature model accuracy\label{subsec:tb_model_performance}} 

\begin{figure}
	\centering
	\small
    \begin{subfigure}[b]{0.49\linewidth}
         \flushright\includegraphics[trim={0cm 1.25cm 0.0cm 0cm},clip,width=\linewidth]{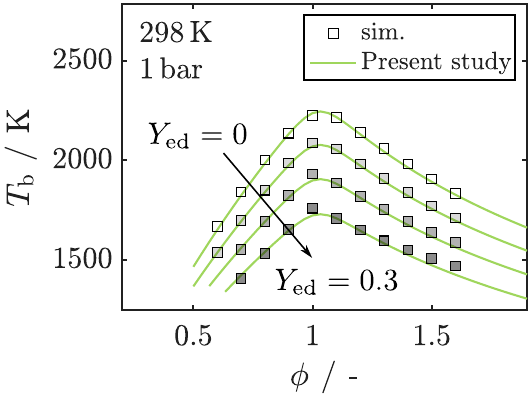}\\
         %
         %
         \flushright\includegraphics[trim={0cm 0.0cm 0.0cm 0cm},clip,width=\linewidth]{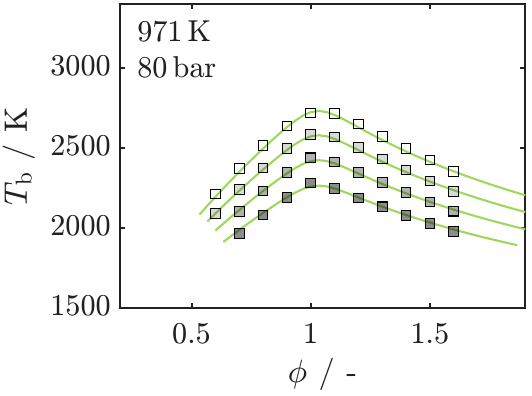}
         \caption{\methane{}/air}
    \end{subfigure}
    \begin{subfigure}[b]{0.49\linewidth}
         \flushright\includegraphics[trim={0cm 1.25cm 0.0cm 0cm},clip,width=\linewidth]{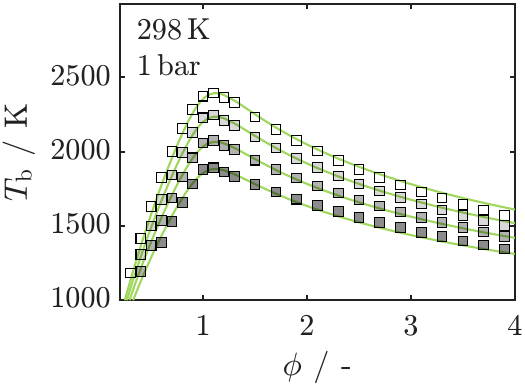}\\
         %
         %
         \flushright\includegraphics[trim={0cm 0.0cm 0.0cm 0cm},clip,width=\linewidth]{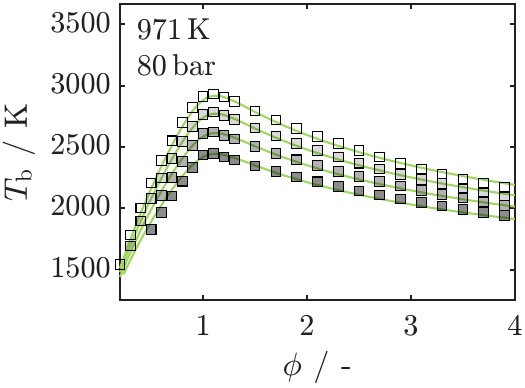}
         \caption{\hydrogen{}/air}
    \end{subfigure}
	\caption{Example adiabatic flame temperature profiles from detailed chemistry (symbols) and the present study's model (lines), for atmospheric burner (top row) and engine (bottom row) representative conditions with varying \(Y_\mathrm{ed} = 0\), 0.1, 0.2, and 0.3 shown in each graph from top to bottom.}
	\label{fig:Tb_vs_phi_absolute}
\end{figure}

The eight-parameter correlation in Eq.~\eqref{eq:core} with the asymmetric shape $S_{\phi_T}$ in Eq.~\eqref{eq:shape_fct} reproduces the detailed-chemistry adiabatic flame temperatures of \methane{}, \hydrogen{}, and \methane/\hydrogen{} (\mbox{$X_{\mathrm{f,H}_2}\!=\!0.4$}, and $0.8$) with high accuracy, yielding $R^2\in[0.993,0.998]$ and mean absolute percentage errors \(f\!<\!1\,\%\) (Table~\ref{tab:phys_params}). Contour maps of the relative deviation over $\phi$, $T_\mathrm{u}$, and $Y_\mathrm{ed}$ and parity plots, provided in the Supplementary Material, show uniformly small errors and a tight clustering along the identity line. Representative $\phi$-sweeps, summarized in Fig.~\ref{fig:Tb_vs_phi_absolute}, confirm that the model reproduces the approximately linear lean-side rise, a smooth single maximum near $\phi_{T,\mathrm{ref}}$, and the fuel-dependent rich-side decay over the full range of $T_\mathrm{u}$, $p$, and $Y_\mathrm{ed}$. Symbols represent the simulation data used to train the correlation. Lines show the correlation results, slightly extrapolated to richer conditions.

\begin{table}[tb]
\centering
\caption{Parameters for the physics-guided $T_\mathrm{b}$ correlation in Eq.~\eqref{eq:core} using the normalized shape $S_{\phi_T}$ from Eq.~\eqref{eq:shape_fct}.}\addvspace{5pt}
\footnotesize
\label{tab:phys_params}
\begin{tabular}{lccccc}
\hline
Parameter & Unit & CH$_4$ & \multicolumn{2}{c}{CH$_4$/H$_2$} & H$_2$ \\
$X_{\mathrm{f,H}_2}$ &[mol/mol] & 0.0 & 0.4 & 0.8 & 1.0 \\
$Y_{\mathrm{f,H}_2}$ &[kg/kg] & 0.0 & 0.077 & 0.335 & 1.0 \\
\hline
$q_T$ &  & 13.72 & 14.43 & 13.47 & 7.369 \\
$A_0$ &  & 2898 & 3005 & 3244 & 3196 \\
$A_T$ &  & -0.396 & -0.397 & -0.397 & -0.327 \\
$\alpha_0$ &  & 0.427 & 0.463 & 0.539 & 0.484 \\
$\tau_Y$ &  & -0.159 & -0.173 & -0.24 & -0.495 \\
$\xi_Y$ &  & 0.966 & 0.963 & 0.967 & 1.063 \\
$m_r$ &  & 0.855 & 0.778 & 0.647 & 0.546 \\
$\phi_{T,\mathrm{ref}}$ &  & 1.027 & 1.028 & 1.039 & 1.062 \\
\hline
Metrics &&&&& \\
\hline
$R^2$ & [-] & 0.998 & 0.998 & 0.997 & 0.993 \\
$f$ & [\%] & 0.446 & 0.444 & 0.497 & 0.971 \\
\hline
\end{tabular}
\end{table}

The model’s parameters are consistent for all fuels and follow expected physical trends. The peak locations $\phi_{T,\mathrm{ref}}$ are close to stoichiometric, with \hydrogen{} showing a slightly richer peak. The negative amplitude $A_T$ indicates that the increase in temperature with preheating becomes smaller, which matches the increased dissociation at higher $T_\mathrm{u}$. The shape parameters $q_T$ and $m_r$ show that \methane{} has a sharper and faster rich-side decay than \hydrogen{}, which fits with the known persistence of high adiabatic temperatures in rich hydrogen flames. The dilution parameter $\xi_Y$ reduces $\Delta T$ as dilution increases. The negative sign of $\tau_Y$ causes this effect to saturate. 

\subsection{Laminar flame speed model \label{subsec:laminar_flame_speed_model}} 

For simple fuels such as \methane{}, asymptotic analysis leads to expressions of the form~\eqref{eq:base_approx}, shown in the Introduction. In these, the inner-layer temperature $T_\mathrm{i}$ and the overall temperature rise $(T_\mathrm{b}-T_\mathrm{u})$ jointly determine both the peak $S_\mathrm{L}$ and its lean/rich flanks. Using these expressions directly would link the peak of $S_\mathrm{L}(\phi)$ to the peak of $T_\mathrm{b}(\phi)$. However, for \hydrogen{}-enriched mixtures, the peaks of $T_\mathrm{b}(\phi)$ and $S_\mathrm{L}(\phi)$ are quite different because of strong transport and chain-branching effects. Making $S_\mathrm{L}$ follow $T_\mathrm{b}(\phi)$ would misrepresent the shape in $\phi$. The following factorization helps address this issue:
\begin{align}
S_\mathrm{L}
&= S_{\text{peak}}(T_\mathrm{u},p,Y_\mathrm{ed})\,\nonumber\\
   &\cdot S_{\phi_S}(\phi;T_\mathrm{u},p,Y_\mathrm{ed})\,
   \cdot \tilde F_{S}(\phi;T_\mathrm{u},p,Y_\mathrm{ed}).
\label{eq:SL_factorization}
\end{align}
The three factors have individual roles: $S_{\text{peak}}$ collects the dependence of $S_\mathrm{L}$ on $T_\mathrm{u}, p$, and $Y_\mathrm{ed}$ at the equivalence ratio of maximum laminar flame speed, $\phi_{S,\mathrm{ref}}$; $S_{\phi_S}$ describes the local variation with $\phi$ around this peak; The normalized flame-structure factor, $\tilde F_S$, reintroduces the influence of the full adiabatic temperature rise and inner-layer structure on the lean and rich flanks without locking the peak to $T_\mathrm{b}(\phi)$, thus, capturing lean and rich flamability limits.

\subsubsection{Peak kinetics model\label{subsec:peak_kinetics_model}} 

The peak kinetics term can be expressed as follows
\begin{align}
S_{\text{peak}} &=
S_0\,
A_p\,
\exp\biggl[-\Theta_a\Bigl(\frac{1}{T_\mathrm{i}}-\frac{1}{T_\mathrm{i,ref}}\Bigr)\biggr]\nonumber \\
&\quad\cdot
\Bigl(\frac{T_\mathrm{u}}{T_\mathrm{i}}\Bigr)^{r_\mathrm{th}}\,
\bigg(\frac{T_\mathrm{b}^\star - \delta T_\mathrm{i}}{T_\mathrm{b}^\star - T_\mathrm{u}}\bigg)^{n}.
\label{eq:SL_peak}
\end{align}
Here, $S_0$ represents a reference speed, $\Theta_a$ is the effective activation temperature, and $\tilde p = \ln(p/p_\mathrm{ref})$ denotes the normalized pressure. The exponent $r_\mathrm{th}$ determines how sensitive the model is to thermal expansion, as measured by the ratio $T_\mathrm{u}/T_\mathrm{i}$. Asymptotic theory suggests that $r_\mathrm{th}$ is close to unity. The Arrhenius-type function mainly describes how $T_\mathrm{i}$ affects the result, with $T_\mathrm{i,ref}$ acting as a reference inner-layer temperature. The last term in Eq.~\eqref{eq:SL_peak} introduces an additional sensitivity of the peak flame speed to the overall adiabatic temperature rise and to the distribution of this rise between the inner layer and the burned gases, as in classical rate-ratio asymptotic theory~\cite{Williams1987asymptotic}.  In this context, $T_\mathrm{b}^\star - T_\mathrm{u}$ is the total flame temperature rise at $\phi_{S,\mathrm{ref}}$, while $T_\mathrm{b}^\star - \delta T_\mathrm{i}$ estimates the part of this rise that happens after the inner layer, with $\delta$ being a small value parameter close to unity. $T_\mathrm{b}^\star$ is obtained from the adiabatic flame temperature model at $\phi_{S,\mathrm{ref}}$~(cf. Sec.~\ref{subsec:adiabatic_flame_temperature_model}).

The inner-layer temperature $T_\mathrm{i}$ in classical asymptotic theory is expressed as $T_\mathrm{i} = -\Theta_\mathrm{i}/\ln(p/B)$~\cite{Goettgens1992_CI_approximation}. To accommodate varying preheat temperatures and dilution levels, we generalize this to
\begin{equation}
T_\mathrm{i}
= \frac{-\Theta_\mathrm{i}}{
    \ln\chi_p
    - B
    + a_T\,(1+\chi_p)\,\tilde T
  }
- a_Y\,Y_\mathrm{ed}.
\label{eq:T0_original}
\end{equation}
$\Theta_\mathrm{i}$ is the activation temperature of the inner reaction zone, $B$ scales the pressure, and $a_T$ controls the sensitivity of the preheating temperature to $T_\mathrm{i,ref}$, where $\tilde T = T_\mathrm{u}/T_\mathrm{i,ref}$. $a_Y$ captures the direct influence of the reduction in $T_\mathrm{i}$ with increasing dilution and reflects the lower amount of reactive mixture.

External dilution also alters collisional and third-body effects, which we capture through an effective pressure
\begin{equation}
p_\mathrm{eff} = p\,(1 + \eta_Y\,Y_\mathrm{ed}),
\label{eq:p_eff}
\end{equation}
and a saturation factor
\begin{equation}
\chi_p = \frac{p_\mathrm{eff}}{p_\mathrm{eff} + p_\mathrm{c}},
\label{eq:chi_p_def}
\end{equation}
where $p_\mathrm{c}$ is a characteristic pressure marking the transition from a low-pressure regime, where $T_\mathrm{i}$ changes strongly with pressure, to a high-pressure regime, where additional pressure or dilution produces only weak changes in $T_\mathrm{i}$.

This structure is consistent with the picture proposed by Han et al.~\cite{Han2024_methane_diluted, Han2024_ammonia_methane}, who showed that external dilution modifies $S_\mathrm{L}$ primarily through changes of the temperature gradient in the inner layer. In our formulation, most of the pressure- and dilution-sensitivity of $S_\mathrm{L}$ is likewise channeled through the modified $T_\mathrm{i}$ in Eq.~\eqref{eq:T0_original}, via the Arrhenius term $\exp[-\Theta_a(1/T_\mathrm{i}-1/T_\mathrm{i,ref})]$, so that $\ln(S_\mathrm{L}/S_{\mathrm{L,ref}})$ is essentially linear in $1/T_\mathrm{i}-1/T_{\mathrm{i,ref}}$. Han et al. write this sensitivity in terms of an undiluted fraction, $Y_\mathrm{ud}$, $\ln(S_\mathrm{L}/S_{\mathrm{L,ref}}) = a\,(1/Y_\mathrm{ud} - 1/Y_{\mathrm{ud,ref}})$, with all operating-point dependence collected in $a(\phi,p,T_\mathrm{u},\text{fuel}, \text{diluent})$. In contrast, we keep $\Theta_a$ nearly fuel-specific and attribute most of the operating-point dependence to $T_\mathrm{i}(p,T_\mathrm{u},Y_\mathrm{ed})$ and to the mild pressure factor~$A_p$, thereby tying the correlation directly to the inner-layer flame structure.

Finally, the pressure factor
\begin{equation}
A_p
= \exp\bigl(-n_p \tilde p - n_{p,2}\tilde p^2\bigr),
\label{eq:Ap_def}
\end{equation}
provides a correction to the pressure dependence already contained in $T_\mathrm{i}$. It is centered at $p_\mathrm{ref}$ such that $A_p(0)=1$. The quadratic term with coefficient $n_{p,2}$ allows for slight curvature, which is particularly important for low- to moderate-pressure \hydrogen/air flames, where $S_\mathrm{L}$ can increase slightly with pressure despite a monotonic decrease in $T_\mathrm{i}$~\cite{Seshadri1994hydrogen}.

The peak kinetics model is parametrized using the subset of simulated data at $\phi_{S,\mathrm{ref}}$. It predicts the simulations with an $R^2$ above $0.999$ and an MAPE of approximately \qty{2}{\%} over $T_\mathrm{u}$, $p$, and $Y_\mathrm{ed}$. These values indicate a robust parameterization with low bias over the operating conditions. Parity and residual plots for $S_\mathrm{peak}$ are provided in the Supplementary Material. The fitted parameters for \methane{}, \hydrogen{}, and the \mbox{$X_{\mathrm{f,H}_2}\!=\!0.4$} and 0.8 blends are listed in Table~\ref{tab:peak_kinetics_par}.

\begin{table}[t]
\centering
\caption{Peak kinetics model parameters and reference conditions for CH$_4$/H$_2$/air mixtures.\label{tab:peak_kinetics_par}}
\footnotesize
\begin{tabular}{lccccc}
\toprule
Parameter & Unit & CH$_4$ & \multicolumn{2}{c}{CH$_4$/H$_2$} & H$_2$ \\
$X_{\mathrm{f,H}_2}$ &[mol/mol] & 0.0 & 0.4 & 0.8 & 1.0 \\
$Y_{\mathrm{f,H}_2}$ &[kg/kg] & 0.0 & 0.077 & 0.335 & 1.0 \\
\midrule
$S_0$ & [cm/s] & 35.42 & 67.88 & 371.48 & 559.57 \\
$\Theta_a$ & [K] & 14783 & 13042 & 10341 & 9396 \\
$n_p$ & [-] & 0.605 & 0.507 & 0.404 & 0.813 \\
$n_{p,2}$ & [-] & -0.051 & -0.048 & -0.022 & -0.006 \\
$r_\mathrm{th}$ & [-] & 0.239 & 0.48 & 0.549 & 1.021 \\
$\Theta_\mathrm{i}$ & [K] & 12952 & 12454 & 13177 & 9204 \\
$B$ & [-] & 8.662 & 8.334 & 8.843 & 6.324 \\
$a_{T}$ & [-] & 1.275 & 1.135 & 1.131 & 0.42 \\
$\delta$ & [-] & 1.072 & 1.074 & 1.091 & 0.997 \\
$n$ & [-] & 2.12 & 2.187 & 2.442 & 2.564 \\
$n_r$ & [-] & 0.317 & 0.618 & 0.295 & 2.75 \\
$a_Y$ & [K] & 570.01 & 470.98 & 209.14 & 0 \\
$\eta$ & [-] & 6.022 & 5.337 & 3.069 & 2.338 \\
$p_{c}$ & [-] & 4.99 & 7.73 & 20.44 & 103.3 \\
\midrule
Reference & & & & &\\
\midrule
$\overline \phi_{S,\mathrm{ref}}$ &  & 1.075 & 1.1 & 1.2 & 1.7 \\
$p_\mathrm{ref}$ & [bar] & 10.0 & 10.0 & 10.0 & 10.0\\
$T_\mathrm{i,ref}$ & [K]   & 1250 & 1250 & 1250 & 1250\\
\bottomrule
\end{tabular}
\end{table}

\subsubsection{Flame-structure effects on flammability limits}

To restore the $\phi$-dependent flame structure effects of classical asymptotics, and thus enable physically sound flammability limits, we introduce a normalized flame structure factor:
\begin{equation}
\tilde F_{S} = 
\left(
\frac{T_\mathrm{b} - \delta T_\mathrm{i}}{T_\mathrm{b} - T_\mathrm{u}}\right)^{n_{r}}
\bigg(\frac{T_\mathrm{b}^\star - \delta T_\mathrm{i}}{T_\mathrm{b}^\star - T_\mathrm{u}}
\bigg)^{-n_{r}},
\label{eq:CTphi_def}
\end{equation}
where $T_\mathrm{b} = T_\mathrm{b}(\phi,T_\mathrm{u},Y_\mathrm{ed})$ and $T_\mathrm{b}^\star = T_\mathrm{b}(\phi_{S,\mathrm{ref}},T_\mathrm{u},Y_\mathrm{ed})$. At the peak equivalence ratio, $\tilde F_S(\phi_{S,\mathrm{ref}})$ is unity, while $\tilde F_S(\phi)$ mainly modifies the lean and rich flanks of the combined $S_\mathrm{L}$-model given by Eq.~\eqref{eq:SL_factorization}. 

As the numerator in Eq.~\eqref{eq:CTphi_def} approaches zero near the flammability limits, $\tilde F_S(\phi)$ also approaches zero and $S_\mathrm{L}$ vanishes. We define $n_{r}$ as a reduced flame-structure exponent. For $n_{r} = n$, it exactly restores the flame structure predicted by asymptotic theory (cf. Eq.~\eqref{eq:base_approx}). By allowing relatively small values with $n_{r} < n$, we successfully reintroduced flammability limits without linking the $T_\mathrm{b}$ shape in fuel-air equivalence ratio too much to $S_\mathrm{L}$.

\subsubsection{$\phi$-shape model\label{subsec:phi_shape_model}} 

The normalised $\phi$-shape function, $S_{\phi_S}$, is constructed as an asymmetric bell curve whose operating point-dependent apex denotes the equivalence ratio with maximum LFS. Therefore, this algebraic function can be easily expressed as a function of the normalized equivalence ratio, $\tilde \phi_S = \ln(\phi/\phi_{S,\mathrm{ref}})$. The model is given by
\begin{align}
S_{\phi_S}
&= M_\mathrm{t}\,2^{1/\gamma}\,
 \cdot \,\exp\bigl(-\beta_r\,\tilde \phi_S\bigr)\nonumber\\
 &\cdot\,\Bigl[1 + \exp\bigl(-( \beta_\ell + \beta_r)\,\gamma\,\tilde{\phi}_S\bigr)\Bigr]^{-1/\gamma},
\label{eq:shape_fct_SL}
\end{align}
with the roundness parameter $\gamma>0$ and lean- and rich-side exponents $\beta_\ell,\beta_{r}>0$. The tilt modifier $M_\mathrm{t}$ allows for slight asymmetries without shifting the peak and is given by
\begin{align}
M_\mathrm{t} &= \exp(\kappa\,h_\mathrm{t}),\\
h_\mathrm{t} &= \tanh(2\tilde{\phi}_{S})\,[1 - \mathrm{sech}(2\tilde{\phi}_{S})],
\end{align}
which can be written in exponential form as
\begin{align}
h_\mathrm{t}
=&\,\Biggl(\frac{2}{1+\exp(-4\tilde{\phi}_{S})}-1\Biggr)
   \Biggl(1-\frac{2\exp(2\tilde{\phi}_{S})}{1+\exp(4\tilde{\phi}_{S})}\Biggr).
\end{align}
In the absence of the tilt term ($\kappa \rightarrow 0$), the shape function in Eq.~\eqref{eq:shape_fct_SL} produces a bell shape in $\ln\phi$ with power-law decay on the lean and rich sides. However, in the simulated training data, we observed a systematic asymmetry in rich \methane{}-dominated mixtures. At elevated pressures and in highly diluted scenarios, the $S_\mathrm{L}$ decrease is dampened on the rich side, so that a symmetrical shape definition is insufficient. The tilt modifier, $M_\mathrm{t}(\tilde\phi_S)$, provides a compact way to account for this so-called "shoulder": fitted values of $\kappa$ remain close to zero in hydrogen and low-pressure/high-temperature cases, and become appreciable only for rich, \methane{}-dominated flames at elevated pressure and dilution. 

\begin{figure}
    \centering
    \small
    \includegraphics[trim={0cm 0cm 0.0cm 0cm},clip,width=\linewidth]{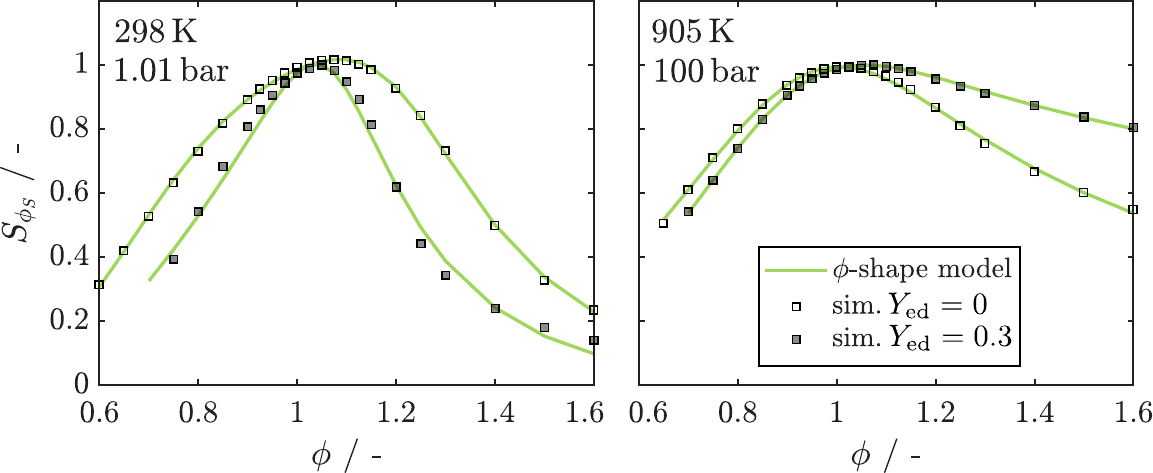}\\
    \makebox[\linewidth][c]{%
        \begin{minipage}{0.8\linewidth} 
            (a) atmospheric \quad\quad\quad\quad\quad (b) engine condition
        \end{minipage}
    }
    \addvspace{5pt}
    \caption{$\phi$-shape model fits for \methane/air mixtures.}
    \label{fig:peak_shape}
\end{figure}

By design, $S_{\phi_S}(0)$ equals one, so that \mbox{$S_{\phi_S}(\phi_{S,\mathrm{ref}})\!=\!1$}. In the lean and rich limits, one recovers power-law behavior,
$S_{\phi_S}\!\propto\!(\phi/\phi_{S,\mathrm{ref}})^{\beta_\ell}$ as $\phi\!\to\!0$ and
$S_{\phi_S}\!\propto\!(\phi/\phi_{S,\mathrm{ref}})^{-\beta_r}$ as $\phi\!\to\!\infty$. Fits for \methane/air mixtures at atmospheric and engine-like conditions, shown in Fig.~\ref{fig:peak_shape}, demonstrate that the model can adapt to the wide variety of $\phi$-shapes observed across pressure, temperature, and dilution. However, individual parameters, $\beta_{\ell},\beta_{r},\gamma$, and $\kappa$, are required for each curve.

To account for operating-point dependence, we express each shape parameter $\psi_j \in \{\beta_{\ell},\beta_{r},\gamma,\kappa\}$ through an affine predictor in normalized pressure, normalized temperature, and external dilution,
\begin{align}
z_{j}
&=
\psi_{j,0} + \psi_{j,p}\,\tilde p 
+ \psi_{j,T}\,\tilde T + \psi_{j,Y}\,Y_\mathrm{ed},
\label{eq:psi_affine}
\end{align}
with coefficients $\psi_{j,k}$ fitted jointly across all mixtures and operating conditions. The physical shape parameters are then obtained via a smooth mapping
\begin{align}
\psi_j = \ln\bigl(1 + \exp(z_j)\bigr),
\label{eq:psi_positive}
\end{align}
so that $\beta_\ell>0$, $\beta_r>0$, $\gamma>0$, and $\kappa\ge 0$ everywhere in the domain. This guarantees strictly positive parameter values needed to capture typical $S_\mathrm{L}(\phi)$ behavior while retaining differentiability with respect to $\tilde p,\tilde T$, and $Y_\mathrm{ed}$.

In addition, we define a peak-shift compensator $\zeta$ with the same affine structure
\begin{equation}
\zeta =
\zeta_{0} + \zeta_{p}\,\tilde p + \zeta_{T}\,\tilde T + \zeta_{Y}\,Y_\mathrm{ed},
\end{equation}
which enters the peak-location relation via
\begin{equation}
\phi_{S,\mathrm{ref}} = \overline{\phi}_{S,\mathrm{ref}}\,\exp(\zeta).\label{eq:phi_peak_S}
\end{equation}
This construction uses the same normalized state variables as the other shape parameters and allows for modest, smooth shifts of the peak location with pressure, preheat temperature, and dilution. It compensates small peak shifts introduced by the asymmetric shape parameters $\beta_\ell$ and $\beta_r$ and keeps the maximum of $S_\mathrm{L}(\phi)$ centered near the observed equivalence ratio of maximum laminar flame speed. In Table~\ref{tab:shape_model_parameters}, the coefficients for $\beta_\ell$, $\beta_r$, $\gamma$, $\kappa$, and $\zeta$ are listed. 

\begin{table}[t]
\centering
\caption{$\phi$-shape model parameters for CH$_4$/H$_2$/air mixtures.\label{tab:shape_model_parameters}}
\footnotesize
\begin{tabular}{lccccc}
\toprule
Parameter & Unit & CH$_4$ & \multicolumn{2}{c}{CH$_4$/H$_2$} & H$_2$ \\
$X_{\mathrm{f,H}_2}$ &[mol/mol] & 0.0 & 0.4 & 0.8 & 1.0 \\
$Y_{\mathrm{f,H}_2}$ &[kg/kg] & 0.0 & 0.077 & 0.335 & 1.0 \\
\midrule
$\beta_{\ell,0}$ & [-] & 2.723 & 1.206 & 4.243 & 8.48 \\
$\beta_{\ell,p}$ & [-] & 1.35 & -0.302 & 0.554 & -0.185 \\
$\beta_{\ell,T}$ & [-] & -10.55 & -6.788 & -4.149 & -0.943 \\
$\beta_{\ell,Y}$ & [-] & 8.998 & -1.784 & 1.845 & -2.547 \\
$\beta_{r,0}$ & [-] & 4.842 & 6.13 & 10.96 & 1.161 \\
$\beta_{r,p}$ & [-] & -1.098 & -2.591 & -0.894 & -0.737 \\
$\beta_{r,T}$ & [-] & -0.802 & 2.61 & -5.219 & 2.584 \\
$\beta_{r,Y}$ & [-] & -4.528 & -12.05 & -5.756 & -2.079 \\
$\zeta_0$ & [-] & -0.024 & 0.153 & 0.122 & -1.234 \\
$\zeta_p$ & [-] & -0.057 & -0.053 & -0.095 & 0.194 \\
$\zeta_T$ & [-] & 0.274 & 0.267 & 0.428 & -1.276 \\
$\zeta_Y$ & [-] & -0.461 & -0.529 & -0.803 & 2.375 \\
$\gamma_0$ & [-] & 4.091 & -0.637 & -2.82 & -2.205 \\
$\gamma_p$ & [-] & 0.911 & 0.192 & -0.186 & 0.368 \\
$\gamma_T$ & [-] & -1.009 & 1.692 & 2.547 & -1.525 \\
$\gamma_Y$ & [-] & 19.04 & 2.849 & -0.141 & 1.495 \\
$\kappa_{t,0}$ & [-] & 3.149 & 2.589 & 4.788 & 0.908 \\
$\kappa_{t,p}$ & [-] & 0.283 & 0.34 & 0.517 & 0.541 \\
$\kappa_{t,T}$ & [-] & -4.119 & -3.954 & -7.721 & -6.679 \\
$\kappa_{t,Y}$ & [-] & 1.77 & 1.167 & 4.564 & 2.951 \\
\bottomrule
\end{tabular}
\end{table}


\subsubsection{Binary blending model\label{sec:blending}} 

For intermediate \hydrogen{} fractions in \methane/\hydrogen/air mixtures, we employ a mass-flux-based blending rule to obtain continuous LFS predictions as a function of blend ratio. Simple mole- or mass-fraction blending and Le Châtelier-type rules, discussed in the introduction, are not used here due to their poor accuracy for \methane/\hydrogen/air. Related empirical blending laws have been proposed, for example, for \hydrogen/\ammonia{} mixtures by Pessina et al.~\cite{PESSINA202225780}, who also observed that per-blend fits are more accurate and recommended their mixing rule mainly as a preliminary screening tool. We treat the present blending model similarly.

For binary \methane/\hydrogen/air, we define the unburned mass fluxes of the neat fuels
\begin{equation}
  \dot{m}_\mathrm{CH_4} = \rho_{\mathrm{u,CH_4}}\,S_{\mathrm{L,CH_4}},\qquad
  \dot{m}_\mathrm{H_2}  = \rho_{\mathrm{u,H_2}}\,S_{\mathrm{L,H_2}},
\end{equation}
where $\rho_{\mathrm{u}}$ is the density of the unburned mixture (determined from the ideal gas law) and $S_{\mathrm{L}}$ is the laminar flame speed at the same operating point $(p,T_\mathrm{u},\phi,Y_\mathrm{ed})$ as the blend. Let $Y_{\mathrm{f,CH_4}}$ and $Y_{\mathrm{f,H_2}}$ denote the fuel-side mass fractions of \methane{} and \hydrogen{} in the blended fuel, with $Y_{\mathrm{f,CH_4}}+Y_{\mathrm{f,H_2}}=1$. The blend mass flux $\dot{m}_\mathrm{bl}$ is modeled as
\begin{equation}
  \dot{m}_\mathrm{bl}^2
  = \frac{c\,Y_{\mathrm{f,CH_4}}\,\dot{m}_\mathrm{CH_4}^2
         +    Y_{\mathrm{f,H_2}}\,\dot{m}_\mathrm{H_2}^2}
         {c\,Y_{\mathrm{f,CH_4}} + Y_{\mathrm{f,H_2}}} \, ,
  \label{eq:mf_blending}
\end{equation}
with \mbox{$\dot{m}_\mathrm{bl}\!=\!\rho_{\mathrm{u,bl}}\,S_{\mathrm{L,bl}}$}. The calibration parameter $c$ controls the curvature of $\dot{m}_\mathrm{bl}^2$ as a function of composition and is calculated for each operating condition, $c(p,T_\mathrm{u},\phi,Y_\mathrm{ed})$. By using the mass fluxes of both neat fuels and a single supporting blend (e.g., at \mbox{$X_{\mathrm{f,H_2}}\!=\!0.8$}), we can solve Eq.~\eqref{eq:mf_blending} for $c$. Once the calibration factor $c$ is known for given $p,T_\mathrm{u},\phi,Y_\mathrm{ed}$, the interpolation in Eq.~\eqref{eq:mf_blending} yields $\dot{m}_\mathrm{bl}^2$ and hence \mbox{$S_{\mathrm{L,bl}}\!=\!\dot{m}_\mathrm{bl}/\rho_{\mathrm{u,bl}}$} between the neat fuels for arbitrary binary blend ratios.

To improve the accuracy of the blending model, we propose a simple extension by introducing an additional support point. Let $c$ depend linearly on the blending ratio,
\begin{equation}
  c = a + b\,X_{\mathrm{f,H_2}}.
\end{equation}
Substituting $c$ into Equation~\eqref{eq:mf_blending} yields a system that can be solved for $a$ and $b$ using the mass fluxes of two supporting blends at the prescribed compositions \mbox{$X_{\mathrm{f,H_2}}\!=\!0.4$} and \mbox{$X_{\mathrm{f,H_2}}\!=\!0.8$}. As intended, this four-point mass-flux-based blending exactly restores the anchor mass fluxes (neat \methane{}, neat \hydrogen{}, and both supporting blends) and yields a smooth interpolation that captures pronounced nonlinearities in the blending behavior.


\section{Results and Discussions\label{sec:results_and_discussions}}

\subsection{Accuracy, extrapolation, and robustness\label{subsec:sl_model_performance}} 
This section evaluates the accuracy and limitations of the proposed laminar flame speed model, including absolute and relative deviations from the training data and extrapolation capability. The model is evaluated using three newly parameterized correlations from the literature and a fully data-driven surrogate model:
\begin{itemize}
    \item Approx-G: the classic approximation formula by Gött-gens et al.\ with the $T_\mathrm{b}$-model from Müller et al.\ and a linear dilution dependence in $Y_\mathrm{ed}$ following Ewald (13 parameters)~\cite{Goettgens1992_CI_approximation, Mueller1997_approximation, Ewald2006_level_flame}, here limited to $\phi \leq \phi_{S,\mathrm{ref}}$,
    \item PowLaw-A: the power-law correlation by Amirante et al.\ (10 parameters)~\cite{Amirante2017}, without $Y_\mathrm{ed}$,
    \item PowLaw-F: the power-law correlation by Harbi and Farooq (36 parameters)~\cite{Harbi2020}, without $Y_\mathrm{ed}$, and
    \item GPR: the model derived from Gaussian process regression described in Sec.~\ref{subsec:fitting_approach}, representing a fully data-driven approach.
\end{itemize}

Neither PowLaw-A nor PowLaw-F contain an explicit external-dilution model. A simple linear scaling in $Y_\mathrm{ed}$ in the spirit of Del~Pecchia et al.~\cite{DelPecchia2020} was tested, but degraded the MAPE by a factor of 3 to 4 for both \methane/air and \hydrogen-containing mixtures. The more advanced log-linear dilution law of Han et al.~\cite{Han2024_methane_diluted, Han2024_ammonia_methane} is not trivially embedded into power-law correlations. Their dilution scaling parameter is pressure-, temperature-, diluent-, and fuel/air-mixture-dependent. This increases the number of parameters and cross-dependencies, which can cause parameter identifiability issues during the fits. For this reason, we restrict PowLaw-A and PowLaw-F to undiluted mixtures ($Y_\mathrm{ed}=0$) and compare dilution effects only for the present model, GPR, and the refitted Approx-G with linear $Y_\mathrm{ed}$ dependence. All coefficients and closed forms are provided in the Supplementary Material.

\begin{figure}
    \centering
    \small
    \includegraphics[trim={0.0cm 0cm 0.0cm 0cm},clip,width=\linewidth]{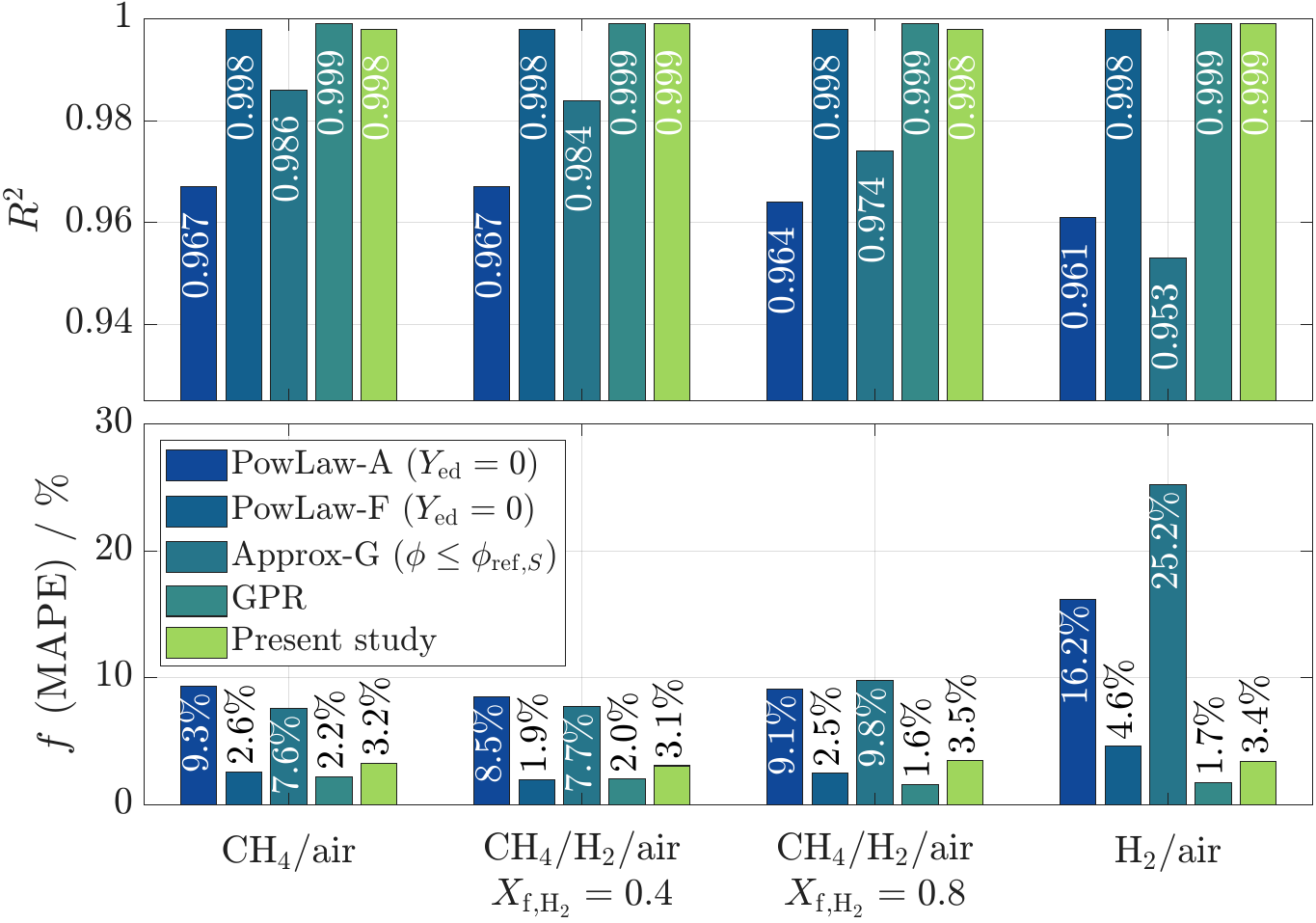} 
    \caption{Comparison of the five LFS models for neat \methane, binary \methane/\hydrogen-blends with $X_{\mathrm{f,H}_2}\!=\!0.4$ and $X_{\mathrm{f,H}_2}\!=\!0.8$, and neat \hydrogen/air mixtures using $R^2$ (top) and MAPE (bottom) metrics; PowLaw-A and PowLaw-F are trained without external dilution, $Y_\mathrm{ed}\!=\!0$, Approx-G is trained without rich conditions, and GPR and the present study are trained against the full simulated matrix, given in Sec.~\ref{subsec:matrix}.}
    \label{fig:SL_approx_benchmark}
\end{figure}

Figure~\ref{fig:SL_approx_benchmark} summarizes the overall $R^2$ and MAPE values of all five models for the three fuel mixtures. Approx-G and PowLaw-A do not reach the target accuracy of $R^2\!>\!0.98$ and $f\!<\!8\,\%$ across all mixtures, as defined by the performance of detailed chemical-kinetic mechanisms (Sec.~\ref{subsec:simulation}). This is attributed to their compactness (13 and 10 parameters, respectively) and to their limited ability to describe the strongly shifted peak and the slowly decaying rich-side flame speeds of \hydrogen{}-containing mixtures. Therefore, PowLaw-A and Approx-G will not be further considered in the model comparison. PowLaw-F, with 36 parameters, satisfies the accuracy requirement for undiluted conditions but does not incorporate external dilution. 
In contrast, both the present physics-guided model and the GPR surrogate achieve high $R^2$ and low MAPE across the full $\phi, T_\mathrm{u}, p$, and $Y_\mathrm{ed}$ space.

\begin{figure}
    \small
    \begin{subfigure}[t]{.543\linewidth}
         \centering
         \includegraphics[trim={0.0cm 0cm 0cm 1.86cm},clip,width=\linewidth]{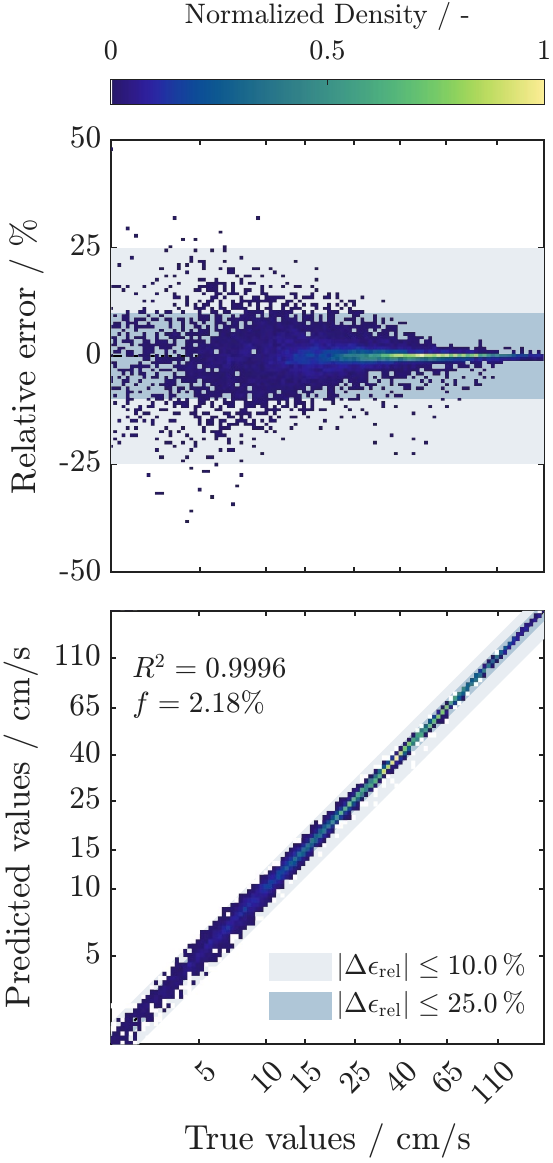}
         \caption{GPR\label{fig:parity_ch4_gpr}}
    \end{subfigure}
    \begin{subfigure}[t]{.445\linewidth}
         \centering
         \includegraphics[trim={1.7cm 0cm 0cm 0cm},clip,width=\linewidth]{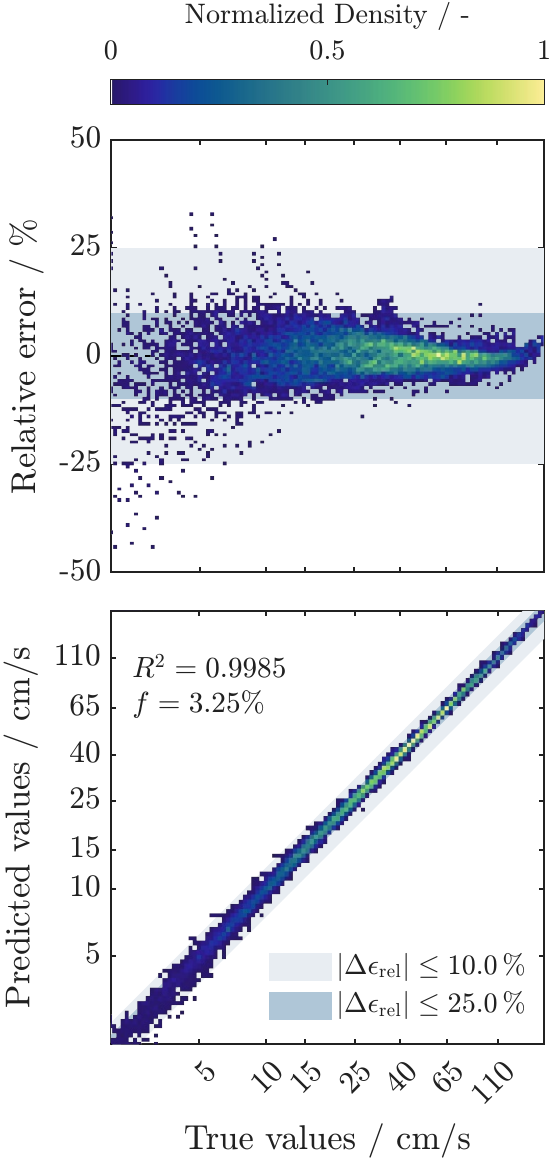}
         \caption{Present study\label{fig:parity_ch4_app_pg}}
    \end{subfigure}
    \caption{Parity diagrams and relative prediction errors of LFS-model predictions for \methane/air mixtures.}
\end{figure}

This is also shown in the parity plots and relative-error distributions for \methane/air, presented for GPR and the present study in Figs.~\ref{fig:parity_ch4_gpr} and \ref{fig:parity_ch4_app_pg}, respectively. Both models produce small deviations that are tightly clustered around the identity line, with nearly symmetric residuals. Furthermore, they show increased deviations for slow LFS, consistent with reduced predictability in less dense regions, such as near the flammability limits. 

\begin{figure*}
	\small
        \begin{subfigure}[b]{0.252\linewidth}
         \flushright\includegraphics[trim={0cm 1.2cm 0.0cm 0cm},clip,width=\linewidth]{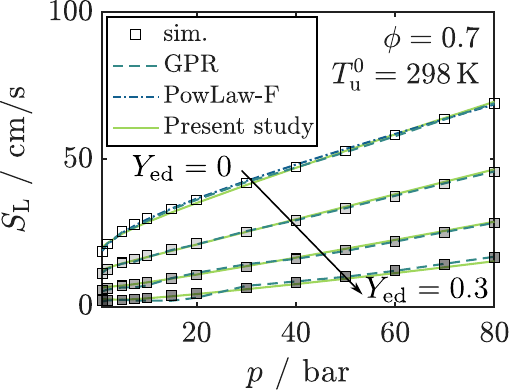}\\
         \flushright\includegraphics[trim={0cm 1.3cm 0.0cm 0cm},clip,width=\linewidth]{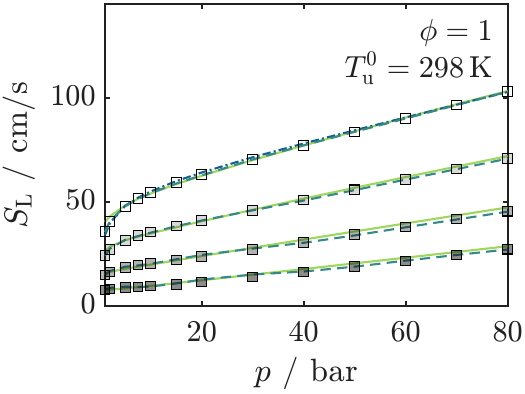}\\
         \flushright\includegraphics[trim={0cm 0.0cm 0.0cm 0cm},clip,width=\linewidth]{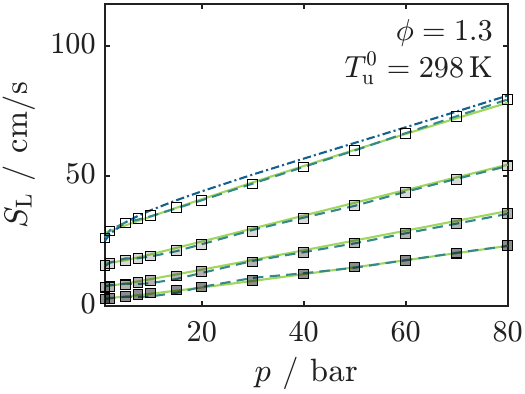}
         \caption{\methane{}/air (isentrope)\label{fig:full_ch4_sl_isentrope}}
    \end{subfigure}
    \begin{subfigure}[b]{0.23\linewidth}
         \flushright\includegraphics[trim={0.6cm 1.3cm 0.0cm 0cm},clip,width=0.98\linewidth]{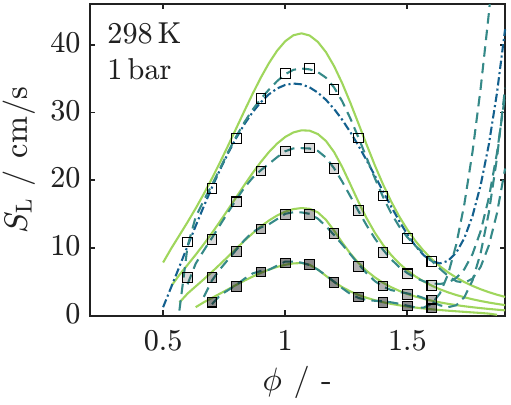}
         \flushright\includegraphics[trim={0.6cm 1.3cm 0.0cm 0cm},clip,width=0.98\linewidth]{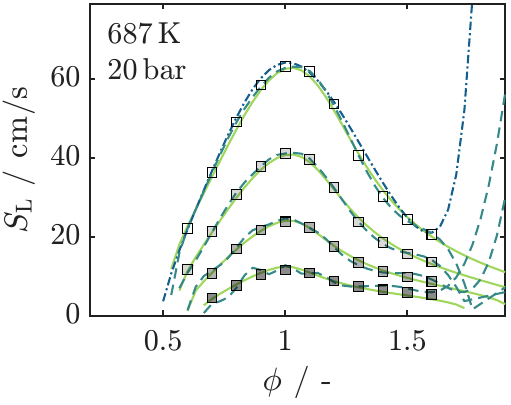}
         \flushright\includegraphics[trim={0.6cm 0.0cm 0.0cm 0cm},clip,width=\linewidth]{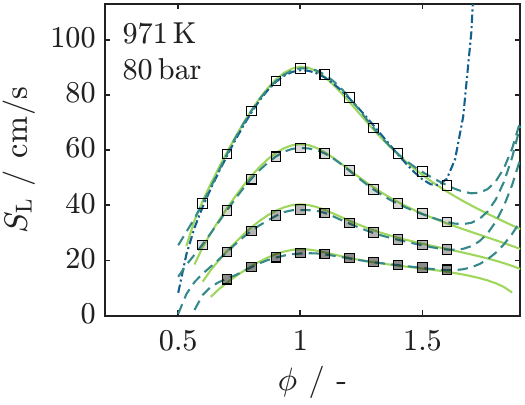}
         \caption{\methane{}/air\label{fig:full_ch4_sl_phi}}
    \end{subfigure}
    \begin{subfigure}[b]{0.23\linewidth}
         \flushright\includegraphics[trim={0.6cm 1.25cm 0.0cm 0cm},clip,width=\linewidth]{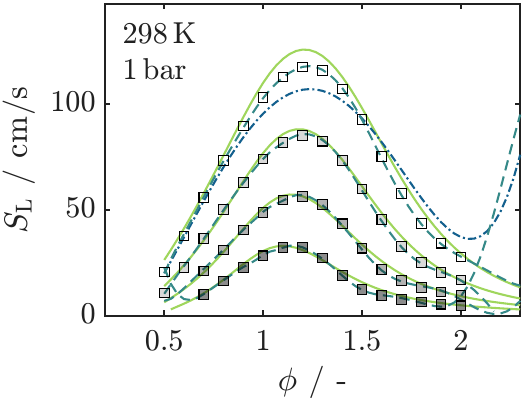}
         \flushright\includegraphics[trim={0.6cm 1.25cm 0.0cm 0cm},clip,width=\linewidth]{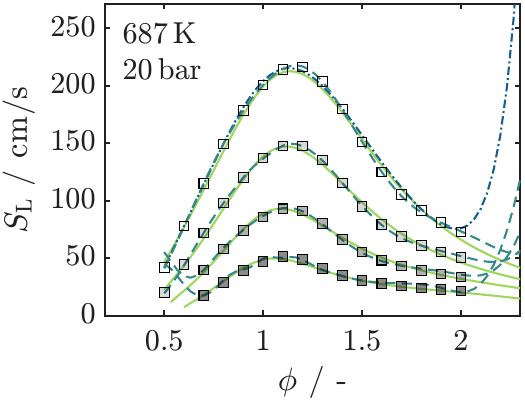}
         \flushright\includegraphics[trim={0.6cm 0.0cm 0.0cm 0cm},clip,width=\linewidth]{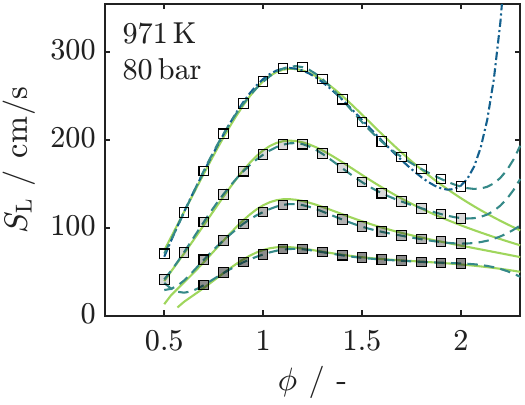}
         \caption{\methane{}/\hydrogen{}/air $X_{\mathrm{f,H}_2}\!=\!0.8$\label{fig:full_blend_sl_phi}}
    \end{subfigure}
    \begin{subfigure}[b]{0.239\linewidth}
         \flushright\includegraphics[trim={0.6cm 1.25cm 0.0cm 0cm},clip,width=0.98\linewidth]{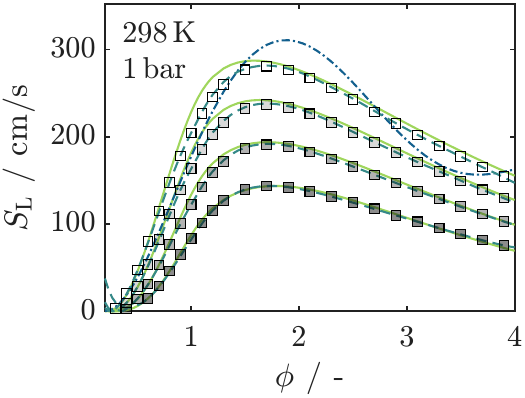}
         \flushright\includegraphics[trim={0.6cm 1.25cm 0.0cm 0cm},clip,width=\linewidth]{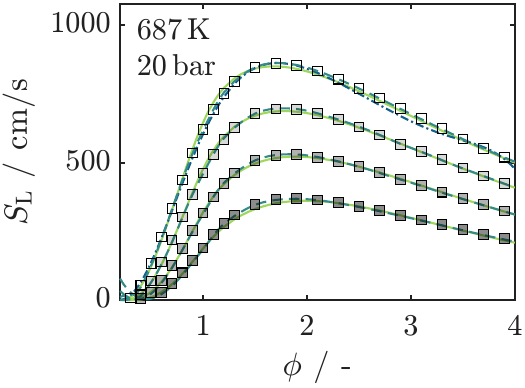}
         \flushright\includegraphics[trim={0.6cm 0.0cm 0.0cm 0cm},clip,width=\linewidth]{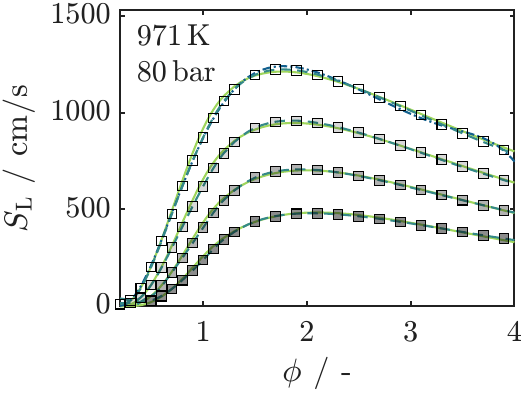}
         \caption{\hydrogen{}/air\label{fig:full_h2_sl_phi}}
    \end{subfigure}
    \\
    \caption{LFS-model results with the complete training data; \(Y_\mathrm{ed} = 0\), 0.1, 0.2, and 0.3 shown in each graph from top to bottom with symbols that increase in opacity.\label{fig:full_sl}}
\end{figure*}

To assess behavior across equivalence ratios, isentropic compression lines, and varying external dilution, Figs.~\ref{fig:full_ch4_sl_isentrope} - \ref{fig:full_h2_sl_phi} compare $S_\mathrm{L}$ from the present model, the reparameterized power-law from Harbi and Farooq (PowLaw-F), and the GPR surrogate against the reference detailed-chemistry simulations for representative burner, gas-turbine, and gas-engine conditions. External dilution varies from \qty{0}{\%} to \qty{30}{\%}, depicted by the gradually increasing opacity of the symbols.     
Isentropic compression is represented for \methane{}/air using a constant isentropic exponent of 1.389 with initial conditions of \mbox{$T_\mathrm{u}^0\!=\!\SI{298}{\kelvin}$} and \mbox{$p^0\!=\!\qty{1.013}{bar}$}. Across all three mixtures, the present model accurately reproduces laminar flame speeds for the isentropic pressure and temperature rise, the lean-side rise, the location and magnitude of the peak, and the rich-side decay, including the dampened decrease in flame speed for rich, \methane{}-dominated flames at elevated pressures and dilution ratios. Under ambient conditions (\qty{1.013}{bar}; \SI{298}{\kelvin}), slight discrepancies occur at the apex and under extremely lean conditions. The dilution-induced attenuation of $S_\mathrm{L}$ and the peak shift and broadening with increasing \hydrogen{}-content are also well captured across the range of conditions. 

The GPR model accurately captures the training data but exhibits uncontrolled extrapolation behavior. This is evident, for example, in the further increase in the laminar flame speed under fuel-rich conditions. PowLaw-F performs well at elevated pressures and temperatures. However, it slightly underestimates ambient conditions, struggles to predict \hydrogen{} LFS, and exhibits unphysical extrapolation behavior under rich conditions. Even though robust extrapolation is typically not the primary design goal when dense training data span the application range, it is important in practice. Correlations are often applied slightly outside their training range or parameterized using limited experimental datasets and then used in broader scenarios. A robust extrapolating model that maintains physically meaningful trends is less prone to overfitting and also exhibits good interpolation capabilities.

\begin{figure*}
	\small
        \begin{subfigure}[b]{0.252\linewidth}
         \flushright\includegraphics[trim={0cm 1.2cm 0.0cm 0cm},clip,width=\linewidth]{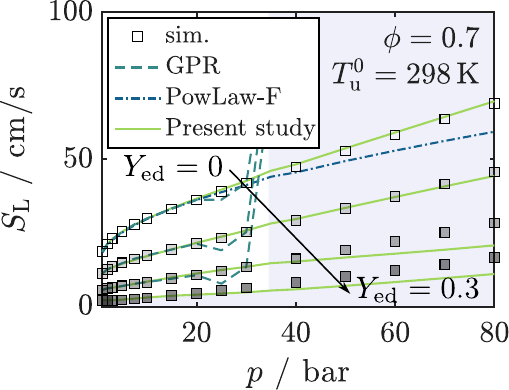}\\
         \flushright\includegraphics[trim={0cm 1.2cm 0.0cm 0cm},clip,width=\linewidth]{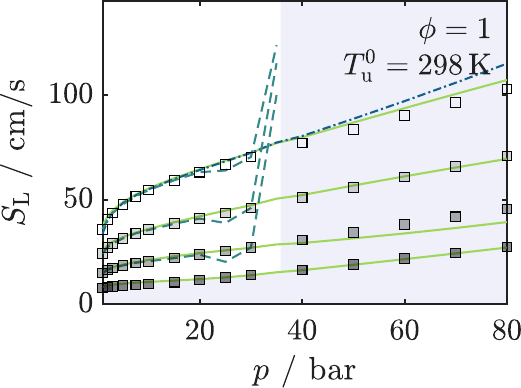}\\
         \flushright\includegraphics[trim={0cm 0.0cm 0.0cm 0cm},clip,width=\linewidth]{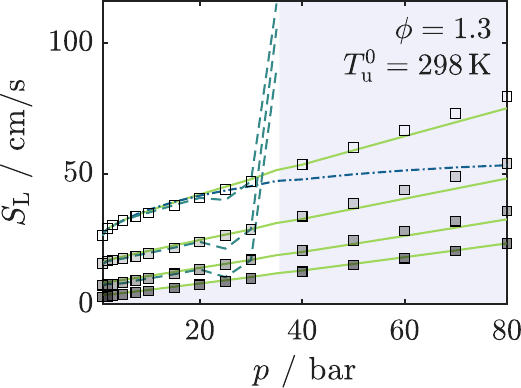}
         \caption{\methane{}/air (isentrope)\label{fig:extrapolate_ch4_sl_isentrope}}
    \end{subfigure}
    \begin{subfigure}[b]{0.23\linewidth}
         \flushright\includegraphics[trim={0.6cm 1.2cm 0.0cm 0cm},clip,width=0.98\linewidth]{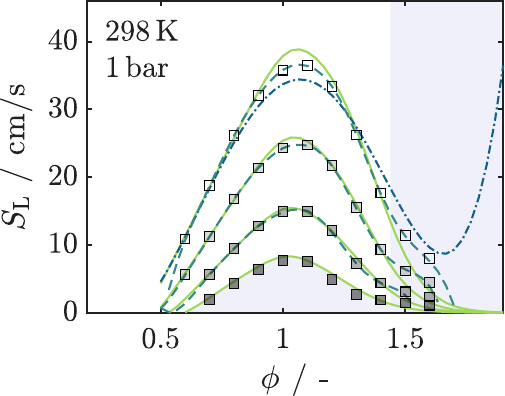}
         \flushright\includegraphics[trim={0.6cm 1.2cm 0.0cm 0cm},clip,width=0.98\linewidth]{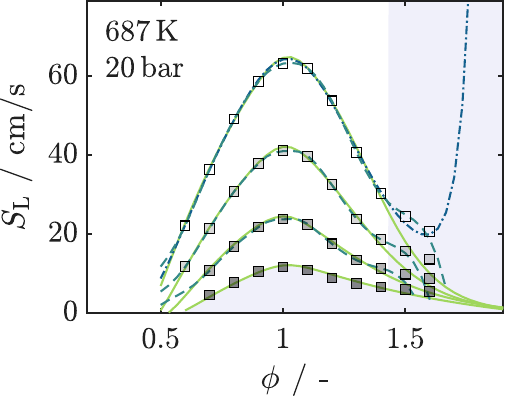}
         \flushright\includegraphics[trim={0.6cm 0.0cm 0.0cm 0cm},clip,width=\linewidth]{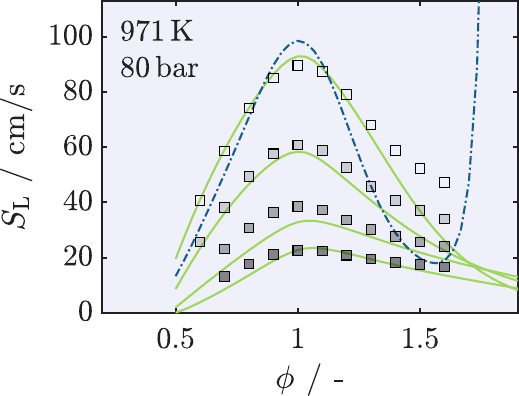}
         \caption{\methane{}/air\label{fig:extrapolate_ch4_sl_phi}}
    \end{subfigure}
    \begin{subfigure}[b]{0.23\linewidth}
         \flushright\includegraphics[trim={0.6cm 1.2cm 0.0cm 0cm},clip,width=\linewidth]{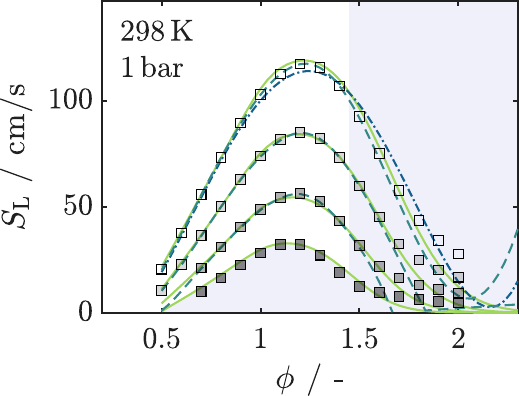}
         \flushright\includegraphics[trim={0.6cm 1.2cm 0.0cm 0cm},clip,width=\linewidth]{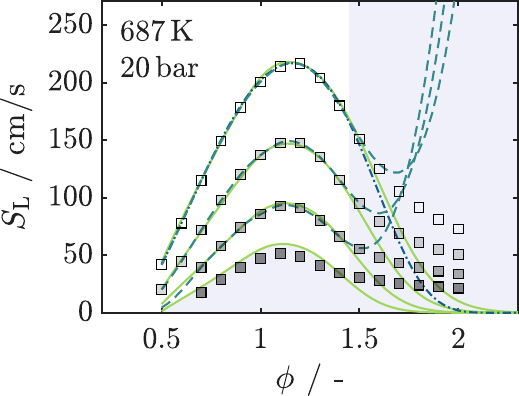}
         \flushright\includegraphics[trim={0.6cm 0.0cm 0.0cm 0cm},clip,width=\linewidth]{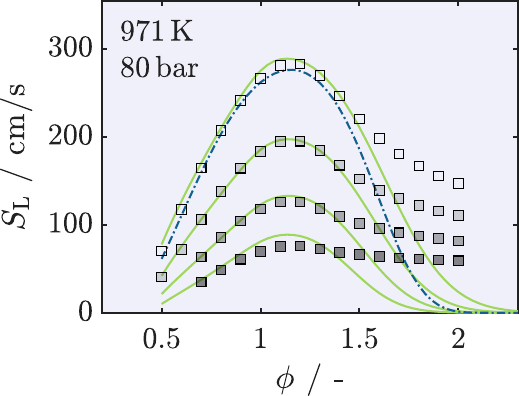}
         \caption{\methane{}/\hydrogen{}/air $X_{\mathrm{f,H}_2}\!=\!0.8$\label{fig:extrapolate_blend_sl_phi}}
    \end{subfigure}
    \begin{subfigure}[b]{0.239\linewidth}
         \flushright\includegraphics[trim={0.6cm 1.2cm 0.0cm 0cm},clip,width=0.98\linewidth]{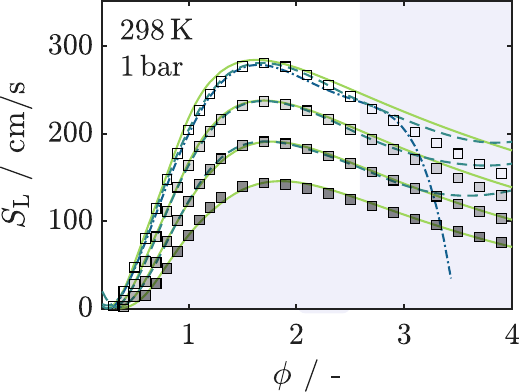}
         \flushright\includegraphics[trim={0.6cm 1.2cm 0.0cm 0cm},clip,width=\linewidth]{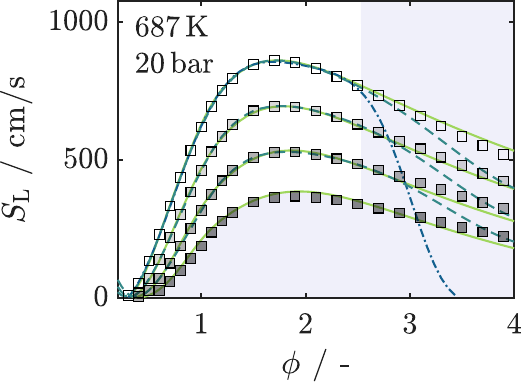}
         \flushright\includegraphics[trim={0.6cm 0.0cm 0.0cm 0cm},clip,width=\linewidth]{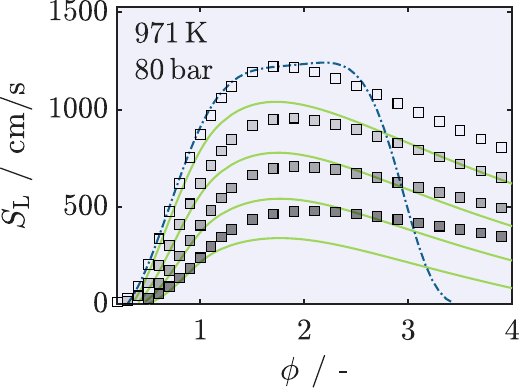}
         \caption{\hydrogen{}/air\label{fig:extrapolate_h2_sl_phi}}
    \end{subfigure}
    \\
    \caption{LFS-model results with training restricted to \(p \in [1.013,30]\,\mathrm{bar}\), \(T_\mathrm{u} \in [298,800]\,\mathrm{K}\), \(Y_\mathrm{ed} \in [0,0.2]\), and \(\phi \in [0.5,1.45]\) for \methane{} and \methane/\hydrogen{} mixtures, and \(\phi \in [0.4,2.5]\) for \hydrogen{}; \(Y_\mathrm{ed} = 0\), 0.1, 0.2, and 0.3 shown in each graph from top to bottom with symbols that increase in opacity; Gray backgrounds denote extrapolated regions.\label{fig:extrapolation_sl}}
\end{figure*}

To assess extrapolation capability, we repeated the training with a deliberately restricted data set, spanning the range of experimentally accessible conditions: $p \in [1.013,30]\,\mathrm{bar}$, $T_\mathrm{u} \in [298,800]\,\mathrm{K}$, $Y_\mathrm{ed}\in[0,0.2]$, and fuel-specific $\phi$ ranges of $\phi\in[0.5,1.45]$ for \methane/\hydrogen/air mixtures and $\phi\in[0.4,2.5]$ for neat \hydrogen/air. Figures~\ref{fig:extrapolate_ch4_sl_isentrope} to \ref{fig:extrapolate_h2_sl_phi} compare the extrapolated $S_\mathrm{L}$ predictions of the different models at the same application-relevant conditions as in Fig.~\ref{fig:full_sl}. The gray backgrounds denote extrapolated regions. The present model maintains an overall good agreement with the detailed-chemistry reference, even in extrapolated pressure and temperature regimes, and with increased external dilution. However, deliberately excluding very rich points from the training data removes information about the rich-side “shoulder”, and the extrapolated model can no longer reproduce this characteristic. Although extrapolation by \qty{50}{bar} into engine-pressure regimes increases deviations in the present model, the trends remain plausible and well-controlled. 

PowLaw-F remains accurate within the training range but develops non-physical curvature and over‑attenuation when extrapolated beyond it. The GPR-derived model is not designed for extrapolation, as evidenced by the isentropic results with arbitrarily increasing and decreasing LFS for \mbox{$p\!>\!\qty{30}{bar}$}. 

\begin{figure}
    \small
    \includegraphics[trim={0.0cm 0cm 0.0cm 0cm},clip,width=\linewidth]{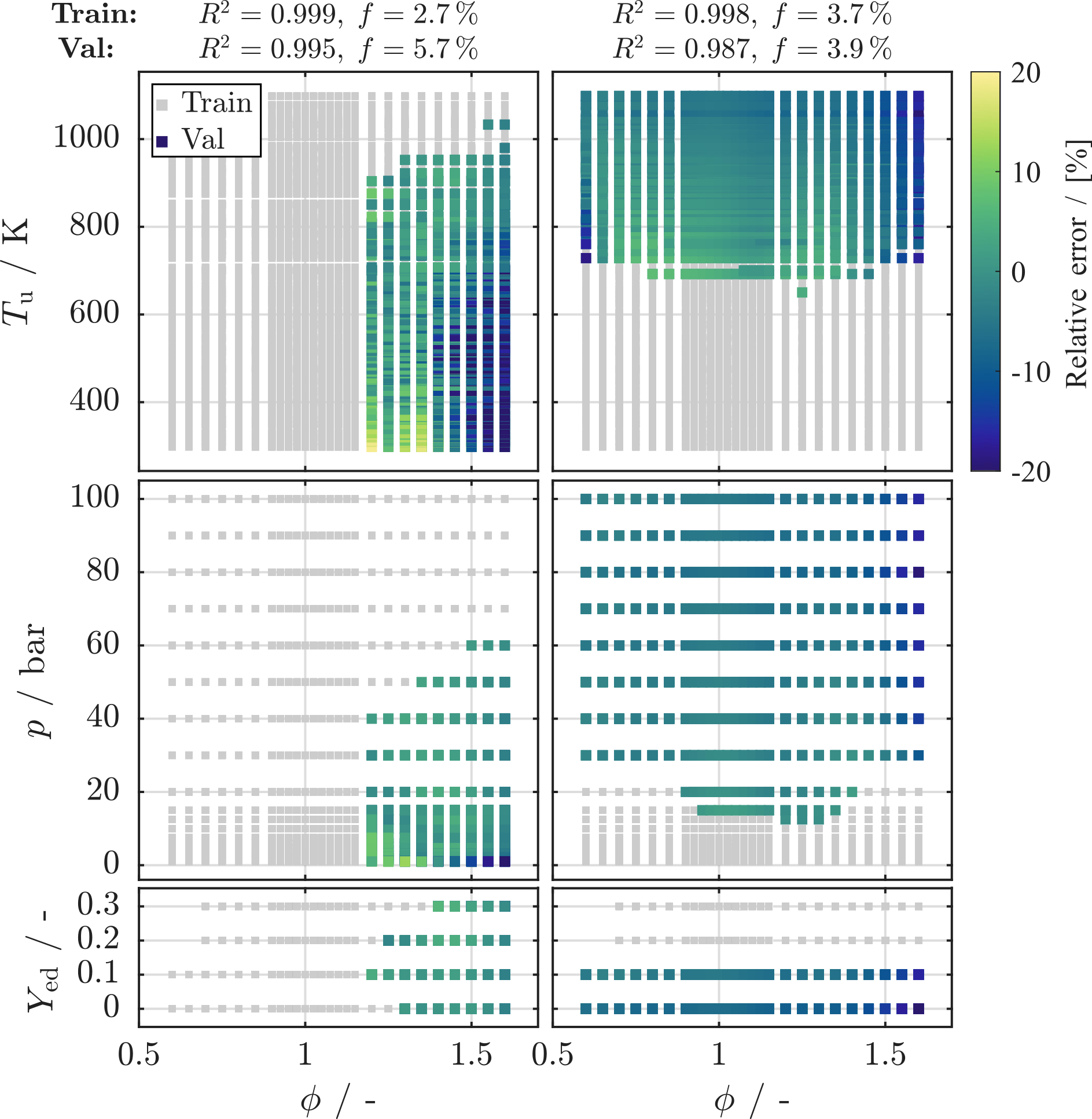} 
    \begin{minipage}[t][][c]{0.37\linewidth}
        (a)
    \end{minipage}
    \begin{minipage}[t][][c]{0.33\linewidth}
        (b)
    \end{minipage}
    \caption{Two example training (gray squares) and validation folds for \methane/air with the colormap representing relative errors of the validation fold; Training and validation \(R^2\) and \(f\) (MAPE) are shown in the figure title.}
    \label{fig:clustered_5fold}
\end{figure}

To quantify robustness with respect to the specific sampling of the training set, we performed a cluster-based 5-fold cross-validation in the four-dimensional space $(\phi,T_\mathrm{u},p,Y_\mathrm{ed})$ for the new correlation, using approximately spherical clusters. For each fold, one cluster was withheld for validation while the remaining clusters were used for training, as shown in Figs.~\ref{fig:clustered_5fold}. Across folds, the present model shows only modest variation in $R^2$ and MAPE between training and validation sets, and the worst-performing folds still remain within the intrinsic uncertainty of the kinetic reference compared to experiments (Sec.~\ref{subsec:simulation}). Folds that predominantly exclude rich conditions~(cf. Fig.~\ref{fig:clustered_5fold}a) lead to larger deviations in the rich regime, as expected, but without qualitative deterioration or unphysical behavior. This indicates that the physics-guided analytical form is not strongly overfit to any particular subset of the training data and that the observed accuracy is robust to moderate changes in the available training data.

Despite overall good agreement, a few limitations of the present model have been previously noted. First, the largest deviations occur near the edges of the $\phi,T_\mathrm{u},p$, and $Y_\mathrm{ed}$ domain. For example, predictions at ambient pressure and temperature and at the highest dilution levels are slightly less accurate than those at intermediate conditions, and restricting the training range to $p \leq 30\,\mathrm{bar}$ slightly improves performance at \qty{1.013}{bar} while mildly degrading it at the highest pressures. This reflects the fact that a single global fit must compromise between different regions of the domain and is most strongly constrained by the densely populated interior; in this context it is worth noting that detailed kinetic mechanisms themselves differ by up to $15\,\%$ under engine-relevant conditions~\cite{Wang2020_methane_correlation}, so modestly increased errors at absolute corner points remain within the intrinsic model uncertainty. Second, the flexible $\phi$-shape model requires relatively complete $\phi$-sweeps, in particular, if the rich behavior must be captured accurately. Otherwise, LFS might decrease more rapidly under super-rich conditions than detailed kinetic models indicate. 

In summary, the resulting deviations remain significantly below~(\mbox{$<\!\qty{4}{\%}$} MAPE) the discrepancies between detailed kinetic mechanisms and experiments~(\mbox{$<\!\qty{8}{\%}$} MAPE). Sporadic larger deviations are observed near the edges of the $\phi,T_\mathrm{u},p$, and $Y_\mathrm{ed}$ domain, where the model is least constrained. The model extrapolates robustly, and physical boundaries, such as flammability limits, are well preserved. Further refinement of the model would be justified only if a closer correspondence with a specific kinetic mechanism were required.

\subsection{Blending model accuracy\label{sec:blending_results}}

\begin{figure}
    \small
    \centering
    \begin{subfigure}[b]{.8\linewidth}
        \includegraphics[trim={0.0cm 0cm -0.5cm 0cm},clip,width=\linewidth]{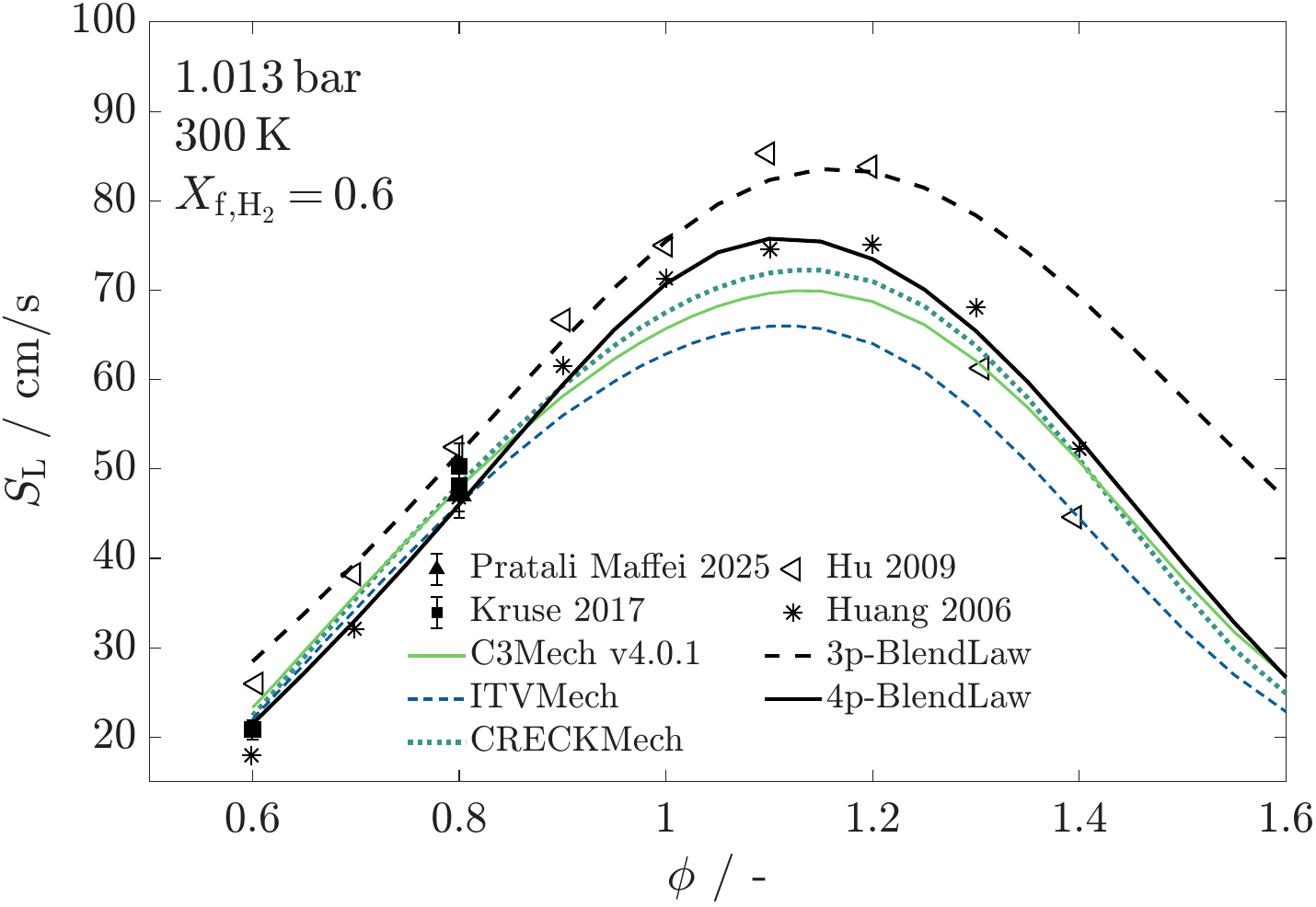} 
        \caption{\methane{}/\hydrogen{}/air at $X_{\mathrm{f,H_2}} = 0.6$}
        \label{fig:SL_CH4H2_pred_0_60bld}
    \end{subfigure}\\
    \begin{subfigure}[b]{.8\linewidth}
        \includegraphics[trim={0.0cm 0cm -0.85cm 0cm},clip,width=\linewidth]{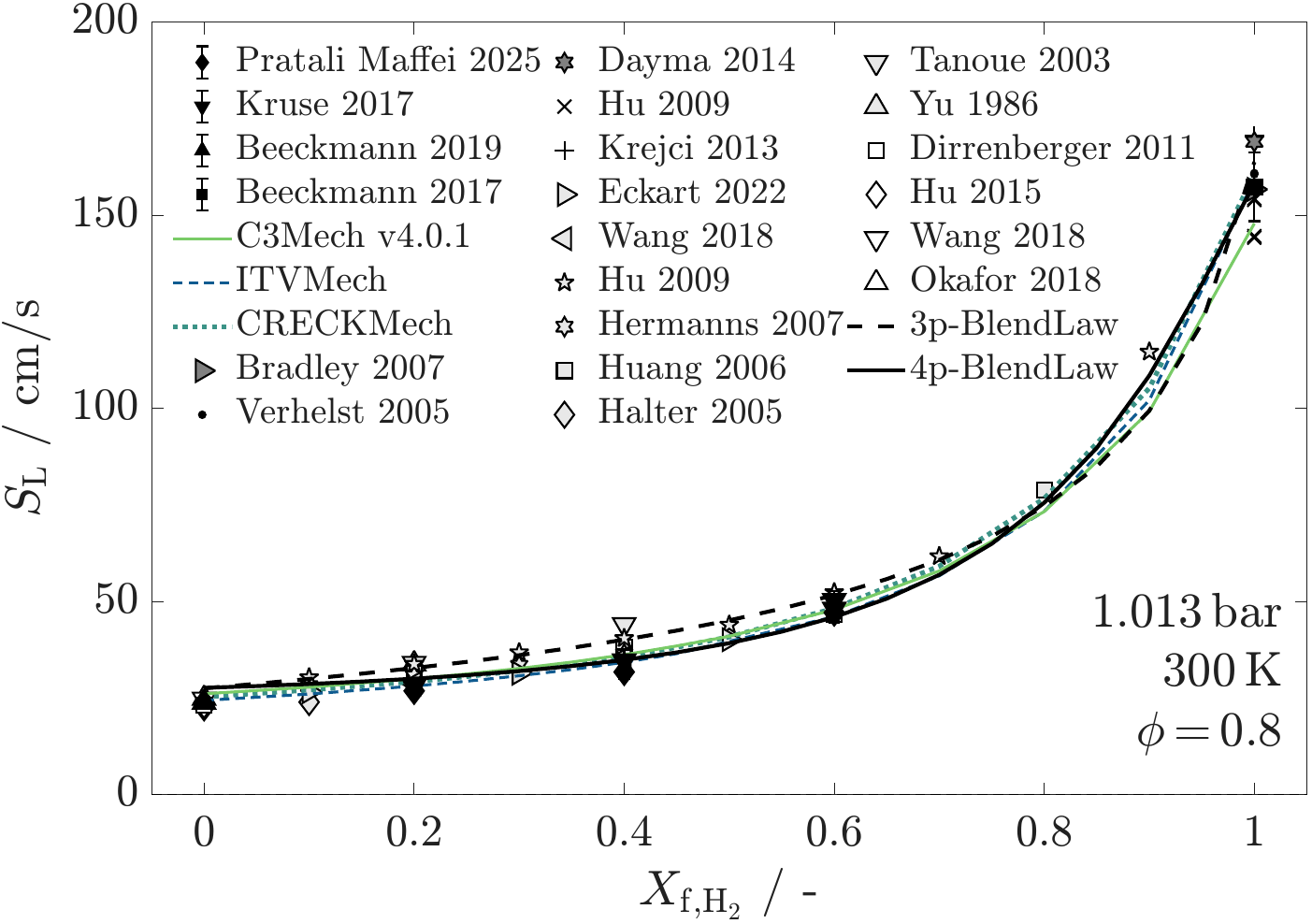}
        \caption{\methane{}/\hydrogen{}/air at $\phi = 0.8$}
        \label{fig:SL_CH4H2__pred_0_80bld_phi08}
    \end{subfigure}\\
    \begin{subfigure}[b]{.8\linewidth}
        \includegraphics[trim={0.0cm 0.0cm -0.85cm 0cm},clip,width=\linewidth]{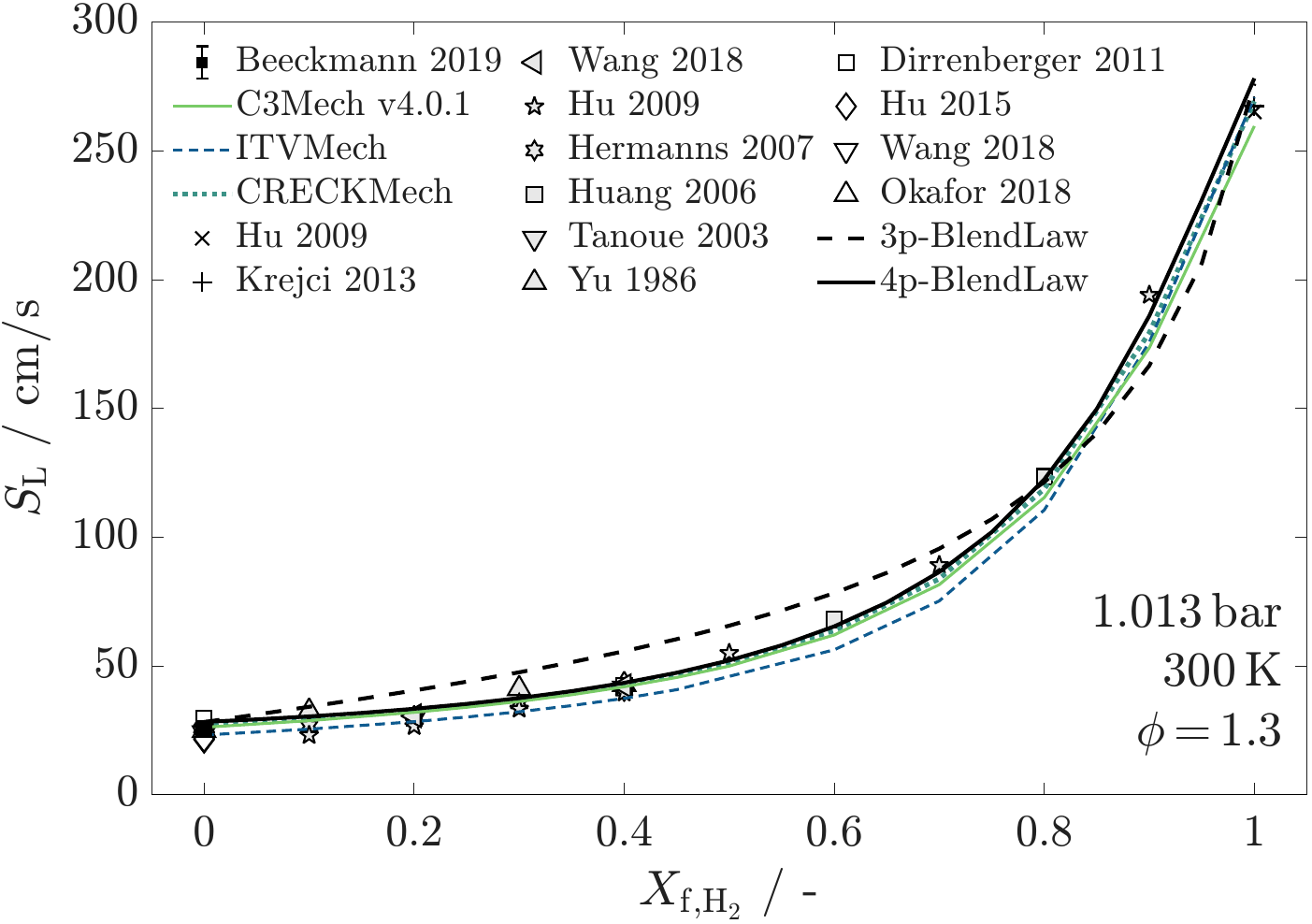}
        \caption{\methane{}/\hydrogen{}/air at $\phi = 1.3$}
        \label{fig:SL_CH4H2__pred_0_80bld_phi13}
    \end{subfigure}\\
    \caption{Laminar flame speed results of the mass-flux-based blending approach for \methane{}/\hydrogen{}/air compared to experimental data from \cite{PrataliMaffei2025C3MechV4, ECM_2017_Kruse_H2_CH4_mix, ICDERS_2017_H2_FlameSpeeds, Beeckmann2019CI_flame_propagation, Hu2009CH4_H2_air_study, Huang2006LBV_H2, Bradley2007hydrogen, Verhelst2005hydrogen, dayma2014peculiar, Krejci2013hydrogen, Eckart2022ch4_h2, Wang2018lbv_ch4_dme_h2, Hermanns2007phd, Halter2005characterization, Tanoue2003JSME, Yu1986_methane_hydrogen, Dirrenberger2011laminar_naturalgas, Okafor2018methaneammonia} and detailed-chemistry simulations.}
\end{figure}

The previous section covered results for $X_{\mathrm{f,H_2}}\!=\!0$, 0.8, and 1.0. Here, we examine how well the mass-flux-based blending rule predicts flame speed for \methane{}/\hydrogen{} mixtures between these points. Figures~\ref{fig:SL_CH4H2_pred_0_60bld} to \ref{fig:SL_CH4H2__pred_0_80bld_phi13} compare \(S_\mathrm{L}\) from experiments, simulations, and the present model for \methane/\hydrogen/air mixtures at ambient conditions (\mbox{$p\!=\!1.013\mathrm{bar}$}, $T_\mathrm{u} = 298\ \mathrm{K}$). We consider both the three-point blending law (3p-BlendLaw), supported by a single blend at $X_{\mathrm{f,H_2}}=0.8$, and the four-point blending law (4p-BlendLaw), supported by blends at $X_{\mathrm{f,H_2}}\!=\!0.4$ and $0.8$. By doing so, the nonlinear blend dependence into three regimes: (1) a weak, nearly linear increase dominated by \methane{}-chemistry~($X_{\mathrm{f,H_2}}\!=\!0$ to 0.4); (2) a nonlinear transition~($X_{\mathrm{f,H_2}}\!=\!0.4$ to 0.8); and (3) a strong, nearly linear increase  dominated by \hydrogen{}-chemistry~($X_{\mathrm{f,H_2}}\!=\!0.8$ to 1.0). The blending-model predictions, along with detailed chemistry simulations and selected experimental data, are used to indicate typical uncertainty in the experiments and predictions.

Figure~\ref{fig:SL_CH4H2_pred_0_60bld} shows a sweep in equivalence ratio at fixed fuel composition, $X_{\mathrm{f,H_2}}\!=\!0.6$. Both mixture variants reproduce the qualitative shape of the laminar flame speed curve and exhibit typical trends across the lean-to-moderately rich range. The 4p-BlendLaw closely follows the C3Mech v4.0.1 reference across the entire $\phi$ range, with deviations that are generally smaller than the spread between different kinetic mechanisms and comparable to the experimental scatter. The 3p-BlendLaw exhibits larger deviations near the composition where no supporting blend is located for all equivalence ratios. 

Figures~\ref{fig:SL_CH4H2__pred_0_80bld_phi08} and \ref{fig:SL_CH4H2__pred_0_80bld_phi13} show the laminar flame speed as a function of $X_{\mathrm{f,H_2}}$ for two representative equivalence ratios, $\phi\!=\!0.8$ and $\phi\!=\!1.3$, respectively. For the fuel-lean case, the 4p-BlendLaw captures the smooth, nonlinear increase of $S_\mathrm{L}$ with hydrogen content and aligns with experiments and simulations across the entire blend range. Differences between the 4p-BlendLaw and the training mechanism remain within, or below, the level of discrepancy between different detailed-chemistry mechanisms and the experimental scatter. The 3p-BlendLaw still provides a reasonable global trend, but shows noticeable deviations in the mid-range compositions, where it lacks a dedicated supporting blend.

For the moderately rich case ($\phi=1.3$), the composition dependence of $S_\mathrm{L}$ becomes more pronounced and strongly nonlinear. The 4p-BlendLaw reproduces the curvature and local slope of simulations and experiments. In contrast, the 3p-BlendLaw tends to under- or over-predict $S_\mathrm{L}$ in intervals that are not closely anchored by supporting blend. However, it still captures the overall monotonic trend.

In summary, these examples demonstrate that the proposed physics-guided LFS correlation, combined with the four-point mass-flux-based blending strategy, yields LFS predictions for \methane/\hydrogen/air mixtures that are competitive with detailed-chemistry simulations. The predictions are consistent with experimental data. Notably, the 4p-BlendLaw maintains predictive accuracy within the bounds of uncertainty inherent to the underlying kinetic model. While the 3p-BlendLaw remains valuable for preliminary assessments, it is more prone to interpolation errors, particularly in blend regions distant from its supporting data points.

\section{Concluding Remarks\label{sec:concluding}}

This work set out to develop and assess a physics-guided analytical correlation for unstretched laminar flame speeds of \methane/\hydrogen/air mixtures with external dilution across \mbox{engine-,} gas-turbine-, and burner-relevant conditions. The main findings can be summarized in terms of the four research questions~(Q1--Q4) posed in the introduction.

First, we estimated the required accuracy for LFS correlations (Q1) by comparing three recent detailed mechanisms with the LFS database. Our database included about 4000 targets from the literature and was extended using new \methane{}/air and \hydrogen{}/air measurements at elevated pressure. The comparison showed that the most advanced kinetic models deviate from the experiments by less than \qty{8}{\%} MAPE, which roughly corresponds to the typical measurement uncertainty. This value sets a realistic accuracy target for LFS correlations. Although the CRECKMech showed slightly better accuracy metrics, we used C3Mech v4.0.1 for our training data generation, as it represents the most recent base chemistry and has the most comprehensive validation approach. 

The new correlation was developed with particular emphasis on an external dilution submodel (Q2). We embedded a similar log-linear dilution dependence, as reported by Han et al.~\cite{Han2024_methane_diluted}, into the LFS model, which is consistent with the Arrhenius-type relations in asymptotic theory. Unlike Han et al., we expressed dilution sensitivity in terms of $T_\mathrm{i}$ instead of an undiluted fraction. The resulting model accurately reproduces the dilution-induced attenuation of $S_\mathrm{L}$ even under fuel-rich conditions. 

Addition of \hydrogen{} to \methane{}/air mixtures was targeted in research question Q3. We showed that the asymmetric $\phi$-shape function for laminar flame speeds successfully captures the shift in maximum point position and width even at high hydrogen substitution rates. Individual parameter sets for each mixture enable maximum accuracy. To integrate a general formulation of blend ratios into the model, we used the mass-flux blending law proposed by Chen et al., which, if constructed with parametrizations of both neat fuels and a supporting blend, delivered accurate trends. By substitution of the model’s calibration factor with a linear function of $X_{\mathrm{f,H}_2}$, we extended the model for two supporting blend ratios, $X_{\mathrm{f,H}_2} = 0.4$ and 0.8. This method provided very accurate predictions and surpasses the model by Chen et al. and other blending formulations.    

Regarding the overall accuracy and robustness of the correlation (Q4), we found that the new model achieves similar accuracy as Gaussian process regression, while remaining differentiable and more robust.  Note that more complex power-law correlations can also provide accurate results. However, a reliable dilution model must still be included, as proposed by Han et al. Robustness of the present model was evaluated through extrapolation tests that deliberately restricted the training data and through cluster-based 5-fold cross-validation. We demonstrated that the model can deliver accurate predictions even under conditions typical of high-load engine operation (e.g., \qty{80}{bar}, \SI{971}{\kelvin}), despite being trained only up to \qty{30}{bar}. Power-law correlations were less predictive, and the GPR failed to predict when extrapolated. The robust extrapolation behavior of our new model indicates that it can be trained solely on measurement data and still provide reliable predictions under application-relevant conditions. 

In conclusion, the proposed correlation is compact, physics-based, fully differentiable, and computationally efficient. It is suitable for CFD, reduced-order combustion models, and real-time control in fuel-flexible, low-carbon combustion systems. The model can be used as a template for other fuel families, such as \ammonia{}/\hydrogen{} mixtures. 

\appendix

\printcredits

\section*{Declaration of competing interest}

The authors declare that they have no known competing financial interests or personal relationships that could have appeared to influence the work reported in this paper.

\section*{Acknowledgments}
We gratefully acknowledge support by the German Research Foundation (DFG) within the DFG Priority Program SPP 2419 HyCam (Grant no. 523874889). Computing time provided at RWTH Aachen University (project number rwth1433) is gratefully acknowledged. The authors also thank Ben Esser, Let\'icia Holtz Telles, and Barnabas Balz for their preliminary work on the data and model analysis.


\bibliographystyle{pci}
\bibliography{ch4_h2}


\small
\baselineskip 10pt


\end{document}